\documentclass[pra,twocolumn, showpacs, superscriptaddress,preprintnumbers,amsmath,amssymb, longbibliography]{revtex4}  % for review and submission

\AtBeginDocument{\usepackage{booktabs}}

\usepackage{graphicx}% Include figure files
\usepackage{bm}% bold math
\usepackage[utf8]{inputenc}
\usepackage{multirow}
\usepackage{makecell}

\usepackage{textcomp}

\usepackage[]{fontenc}

\usepackage{amsmath}
\usepackage{physics}

\usepackage{hyperref}

\begin{document}

%\preprint{APS/123-QED}

\title{mm-wave Rydberg-Rydberg resonances as a witness of intermolecular coupling in the arrested relaxation of a molecular ultracold plasma}% Force line breaks with \\

\author{R. Wang}
\affiliation{Department of Physics \& Astronomy, University of British Columbia, Vancouver, BC V6T 1Z3, Canada}
\author{J. Sous}  
\affiliation{Department of Physics, Columbia University, New York, NY 10027, USA}
\author{M. Aghigh} 
\affiliation{Department of Chemistry, University of British Columbia, Vancouver, BC V6T 1Z3, Canada}
\author{K.  L. Marroqu\'in} 
\affiliation{Department of Chemistry, University of British Columbia, Vancouver, BC V6T 1Z3, Canada}
\author{K. M. Grant} 
\affiliation{Department of Chemistry, University of British Columbia, Vancouver, BC V6T 1Z3, Canada}
\author{F. B. V. Martins} 
\altaffiliation[Present address: ] {Department of Chemistry and Applied Biosciences, ETH Zurich, 8093 Zurich, Switzerland}
 \affiliation{Department of Chemistry, University of British Columbia, Vancouver, BC V6T 1Z3, Canada}
 \affiliation {Department of Chemistry, Kenyon College, Gambier, Ohio 43022 USA}
\author{J. S. Keller}
 \affiliation{Department of Chemistry, University of British Columbia, Vancouver, BC V6T 1Z3, Canada}
 \affiliation {Department of Chemistry, Kenyon College, Gambier, Ohio 43022 USA}
\author{E. R. Grant}
\email[Author to whom correspondence should be addressed. Electronic mail:  ]
{edgrant@chem.ubc.ca}
\affiliation{Department of Physics \& Astronomy, University of British Columbia, Vancouver, BC V6T 1Z3, Canada}
\affiliation{Department of Chemistry, University of British Columbia, Vancouver, BC V6T 1Z3, Canada}

\begin{abstract}
Out-of-equilibrium, strong correlation in a many-body system can trigger emergent properties that act in important ways to constrain the natural dissipation of energy and matter.  Networks of atoms, intricately engineered to arrange positions and tune interaction energies, exhibit striking dynamics.  But, strong correlation itself can also act to restrict available phase space.  Relaxation confined by strong correlation gives rise to scale invariance and density distributions characteristic of self-organized criticality.  For some time, we have observed signs of self-organization in the avalanche, bifurcation and quench of a state-selected Rydberg gas of nitric oxide to form an ultracold, strongly correlated ultracold plasma.  The robust arrested relaxation of this system forms a disordered state with quantum-mechanical properties that appear to support a coherent destruction of transport.  Work reported here focuses on initial stages of avalanche and quench, using the mm-wave spectroscopy of an embedded quantum probe to characterize the intermolecular interaction dynamics associated with the evolution to plasma.  Double-resonance excitation prepares a Rydberg gas of nitric oxide composed of a single selected state, $n_0f(2)$ (referring to a particular principal quantum number from $n_0 = 30$ to 70 in the $f$ ($\ell=3$) Rydberg series converging to the NO$^+$ rotational state, $N^+=2$).  Penning ionization, followed by an avalanche of electron-Rydberg collisions, forms a plasma of NO$^+$ ions and weakly bound electrons, in which a residual population of $n_0$ Rydberg molecules evolves to high-$\ell$.  Predissociation depletes the plasma of low-$\ell$ molecules.  Relaxation ceases and $n_0 \ell(2)$ molecules with $\ell \ge 4$ persist for very long times, evidently under collision-free conditions.  At short times, excitation spectra of mm-wave Rydberg-Rydberg transitions, $n_0f(2) \rightarrow (n_0 \pm 1)g(2)$ mark the rate of electron-collisional $\ell$-mixing.  At long times, $n_0\ell(2) \rightarrow (n_0 \pm 1)d(2)$ depletion resonances signal collision-free energy redistribution in the basis of central-field Rydberg states.  The widths and asymmetries of Fano lineshapes witness the degree to which coupling to the arrested bath broadens the bright state as well as how bright-state predissociation mixes the network of levels in the localized ensemble.   

\end{abstract}


\maketitle

\newcommand{\omegaone}{$\omega_1$ }
\newcommand{\omegatwo}{$\omega_2$ }
\newcommand{\us}{$\mu$s}
\newcommand{\Gone}{G$_1~$}
\newcommand{\Gtwo}{G$_2~$}
\newcommand{\Gthree}{G$_3~$}
\newcommand{\cm}{cm$^{-1}$}

%%%%%%%%%%%%%%%%%%%%%%%%%%%%%%%%%%%%%%%%%%%%%%%%%%%%%%%%%%%%%%%%%%%%%%%%%%%%%%%%%%%%%%%%%%%%%%%%%%%%%%

\section{Introduction}

Strong correlation operates in many-body systems out of equilibrium to constrain the mobility of matter and energy.  Local potentials oppose ergodic driving forces, with far-reaching dynamical implications.  Diverse emergent phenomena, including glassiness \cite{markland2011quantum,olmos2014out,everest2016emergent}, topological phases \cite{moore2013theory,lao2018classical,de2019observation}, quantum magnetism \cite{glaetzle2014quantum,zeiher2016many,harris2018phase,koepsell2019imaging}, fractional quantum Hall states \cite{chandran2016eigenstate,chen2018does}, and high-temperature superconductivity \cite{mitrano2016possible,giannetti2016ultrafast,basov2017towards,tokura2017emergent,kennes2017transient}, all owe defining characteristics to constrained or directed transport either in the classical or quantum regime. 

Quantum systems quenched to a state of large disorder can exhibit coherent destruction of transport, challenging conventional notions of statistical thermodynamics \cite{eisert2015quantum,d2016quantum,gogolin2016equilibration}.  Entangled particle degrees of freedom give rise to local integrals of motion \cite{chandran2015constructing,imbrie2017local,de2017stability,abanin2019colloquium}, which operate to govern energy utilization, and drive periodic or directed conductivity in a limited configuration space \cite{nandkishore2014spectral,swingle2017slow,rubio2019many}, and give rise to many-body localization \cite{yao2014many,nandkishore2015many,lev2015absence,choi2016exploring}.    

Such evident effects of strong correlation lead naturally to the question of whether correlation itself can drive the dynamics necessary to quench a homogeneous system to a state with large disorder far from equilibrium.  If so, can spectroscopic means, developed with model systems in mind \cite{goihl2019experimentally,pagliero2020optically}, serve to probe constrained evolution in general, and yield signatures of the microscopic interactions that govern underlying dynamics ?  

These questions call for the experimental study of systems with correlations strong enough to drive out-of-equilibrium self-assembly.  By construction, such a system ought to afford the simplicity of a quantum-state-selective response that signals the characteristic underlying particle-particle interactions.  An effective measure should show whether strong coupling can drive out-of-equilibrium quantum evolution from homogenous initial conditions to a disordered landscape.  Experimental study along these lines could well confirm conditions under which correlated network interactions support a quench to an ensemble best described in terms of many-body localization, signified by locally conserved quantities and suppressed conductivity.  

Rydberg gases offer particular advantages as materials from which to form out-of-equilibrium strongly correlated many-body systems.  Rydberg-Rydberg interactions occur with coupling energies typical of condensed matter in systems with the density of a rarified gas \cite{browaeys2016experimental}.   Cooperative behaviour in atomic Rydberg ensembles ranges from precisely defined pairwise and higher-order coherent phenomenon \cite{vsibalic2016driven,firstenberg2016nonlinear,gambetta2020engineering,browaeys2020many} to aggregation \cite{garttner2013dynamic,Schempp2014,urvoy2015strongly}, dissipation \cite{lesanovsky2014out,goldschmidt2016anomalous,letscher2017bistability,bernien2017probing}, non-equilibrium phase transitions \cite{carr2013nonequilibrium,malossi2014full} and avalanche to plasma \cite{Killian2007,weller2019interplay,review}.  Recent experiments have established evidence for self-organizing criticality in ultracold and room-temperature atomic gases \cite{helmrich2020signatures,ding2020phase}.   

Here, we investigate the quantum mechanical properties of the out-of-equilibrium state formed when a Rydberg gas of nitric oxide entrained at milli-Kelvin temperature in a skimmed supersonic molecular beam evolves via a sequence of electron-impact avalanche, bifurcation and quench to form a strongly coupled, ultracold plasma \cite{Plasma_prl,Sadeghi:2014,schulz2016evolution,Haenel2017,Haenel.2018}.  

We take information from experimental observations to explain this state of arrested relaxation as a self-assembled, spatially correlated quantum system of randomly interacting dipoles of random energies \cite{Sous2018,Sous2019}.  These dipoles populate a wide distribution of Rydberg and excitonic states concentrated in an energy interval within a few hundred GHz of the ionization threshold, doped by a trace population of more deeply bound Rydberg molecules distributed over principal quantum numbers ranging from 30 to 80.  

The arrested phase evolves to approximately the same plasma density, regardless of the initial principal quantum number and density of the Rydberg gas \cite{Schulz-Weiling2016,review}.  But, evidence shows that this self-organization relies on the strong correlation created by a wave of locally balanced numbers of NO$^+$ ions and Rydberg molecules \cite{review}.  The yield of stable plasma thus depends very sensitively on the survival of the selected Rydberg molecules as determined by their predissociative lifetimes.  

Among the exclusively $N=1$ levels accessible at a given initial principal quantum number, $n_0$, only states in the orbital angular momentum, $\ell=3$ series converging to $N^+=2$ $\left(n_0f(2) \right)$ live long enough to support evolution to a long-lived plasma state.  Higher-$\ell$, $n_0g(2)$ states live an order of magnitude longer \cite{Vrakking1995,Bixon,Remacle1998,Murgu2001}, and we find that mm-wave radiation tuned to stimulate transitions from an allowed $n_0f(2)$ level to adjacent states of higher angular momentum, $(n_0 \pm 1)g(2)$, substantially increases the density of plasma that survives the first 300 ns of avalanche and persists to form a late-peak electron signal after 20 $\mu$s or more \cite{mmWave1}. 

A mm-wave field tuned to drive the overlapping transitions from $n_0f (2)$, $n_0g (2)$ and $n_0\ell (2)$ to neighbouring $(n_0 \pm 1)d(2)$ states drops the angular-momentum barrier, permitting nanosecond predissociation \cite{Vrakking1995}.  During the first 300 ns of plasma evolution, avalanche electrons collide with Rydberg molecules.  This scrambles the orbital angular momentum of photo-populated $\ell = 2$ levels as they form, which prevents early mm-wave transitions to $nd$ states from depleting the Rydberg population.  

However, after one or two microseconds, we find that electron collisions cease, evidenced by the fact that a transition to $\ell=2$ produces a resonant dip in the plasma signal.  At its centre frequency, this transition suppresses the plasma signal to an amplitude near zero, even though delayed selective field ionization shows that the distribution of states evolves by then to contain fewer than one percent in the $n_0$ level resonant with the mm-wave field.  Thus, the $\omega_{\rm mm}$ excitation spectrum itself shows direct evidence that collisionless energy transfer must occur to fill the hole burned in the distribution of Rydberg states.  

The excitation spectrum of Rydberg-Rydberg transitions to bright states of both higher and lower orbital angular momentum displays Fano lineshapes.  These asymmetric features are characteristically broadened and displaced by coupling that redistributes spectral weight to the localized network of coupled Rydberg levels that forms an underlying quasi-continuum of dark states.  

Reading the nature of this coupling directly in the linewidth, $\Gamma$ and asymmetry parameter, $q$, these spectra show direct evidence that predissociation, which transfers population from the spectroscopically active $(n_0 \pm 1)d(2)$ states to the open-system of ${\rm N(^4S) + O(^3P)}$ atoms, bridges the closed plasma ensemble to a thermal continuum.  This evidently strengthens coupling of the bright state to the localized plasma bath and acts to turn up energy mobility in the strongly coupled out-of-equilibrium system.   

Thus, in the nitric oxide molecular ultracold plasma we realize an out-of-equilibrium many-body system in which strong correlation drives self-organization to a canonical state of arrested relaxation, in which the mm-wave excitation spectrum of an embedded quantum probe gauges the degree to which the population of a particular state quenches or favours ergodicity.  

\section{Experimental}

\subsection{Formation of Rydberg gas and its evolution to ultracold plasma}

\begin{figure}[h!]
%\centering
\includegraphics[scale=0.55]{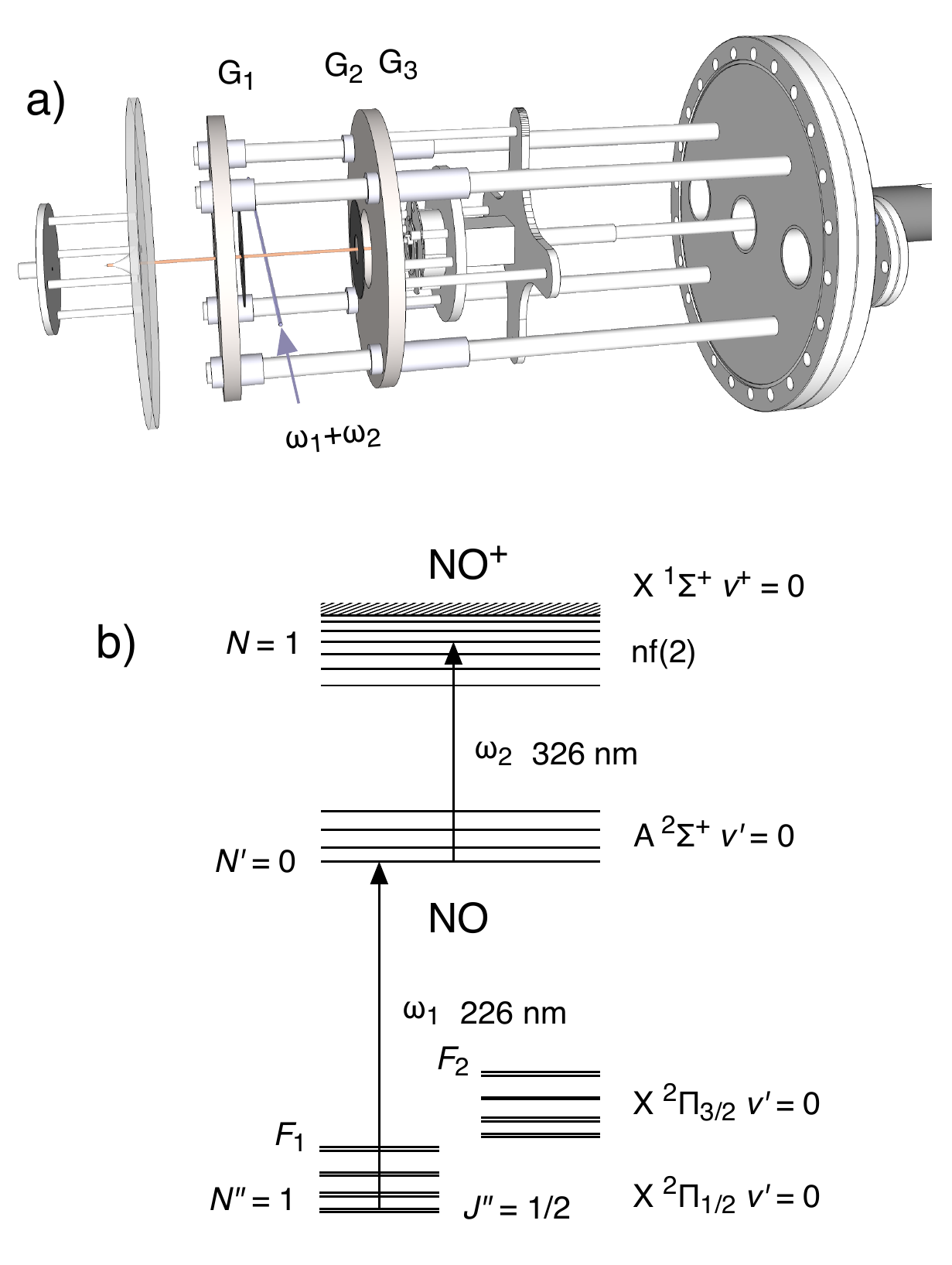}
\caption{(a) Co-propagating laser beams, $\omega_1$ and $\omega_2$, cross a molecular beam of nitric oxide between entrance aperture G$_1$ and grid G$_2$ of a differentially-pumped vacuum chamber. (b)  Nd:YAG-pumped dye lasers ($\omega_1$ and $\omega_2$)  pump ground state NO first to the excited $A\ ^2 \Sigma^+ \ N^\prime=0$ state and then to a Rydberg level with $N=1$. }
\label{fig:diagram}
\end{figure}

Nitric oxide gas, entrained 1:10 in a pulsed supersonic jet of helium, expands from a stagnation pressure of 5 bar. This jet travels 35 mm to pass through a 1 mm diameter skimmer separating the source and experimental chambers of a differentially pumped vacuum system, as diagrammed in Figure~\ref{fig:diagram}a.  A pair of co-propagating Nd:YAG pumped dye laser pulses ($\omega_1 + \omega_2$) intercepts the NO molecular beam between two grids G$_1$ and G$_2$ to produce a state-selected Rydberg gas.  Normally, we hold  G$_1$ to ground and adjust G$_2$ over a range of $\pm 100$ mV to create a field-free region between G$_1$ and G$_2$ during laser excitation.  

The $\omega_1$ pulse, tuned to 226 nm and adjusted to an output energy between 1 and 10 $\mu$J, excites ground state NO molecules to the rotationally selected electronically excited intermediate state, A $^2\Sigma^{+} $ $N^{'}=0$.  A second dye laser pulse, $\omega_2$, tuned from 327 nm to 330 nm, promotes these selected  A $^2\Sigma^{+}$ molecules to high Rydberg states with electron binding energies in the range from 1 to 120 cm$^{-1}$ below the $N^+=0$ ionization limit of NO$^+$ (Figure \ref{fig:diagram}b). 

Parity and angular momentum selection rules from intermediate $N^{'}=0$ confine Rydberg states populated by double resonance to those with total angular momentum neglecting spin of $N=1$. Among optically accessible Rydberg states, only those of principal quantum number $n$ in the $f$ series ($l=3$) converging to the $N^+=2$ limit, $nf(2)$ for $n$ from 30 to 80, have sufficient lifetime to form a Rydberg gas that evolves to form an ultracold plasma.

Normally, we saturate the $\omega_2$ transition, so the intensity of $\omega_1$ determines the initial density of the Rydberg gas.  A $\omega_1$ pulse energy that exceeds 8 $\mu$J saturates the X to A transition, which yields an upper-limiting initial Rydberg gas density in excess of $10^{12}$ cm$^{-3}$ \cite{Plasma_prl,schulz2016evolution}.  At any fixed $\omega_1$  intensity, the radiative lifetime of the A-state affords a precise means to reduce the initial Rydberg gas density by choosing a nonzero $\omega_1-\omega_2$ delay \cite{Settersten2009Radiative}. 

For the purposes of the present experiment, we set $\omega_1$  intensity to equal 2 $\mu$J per pulse and adjust the  $\omega_1 -\omega_2$ delay to a value of 500 ns, which forms a Rydberg gas with an initial density of about $10^{11}$ cm$^{-3}$.  Prompt Penning ionization in the leading edge of the Rydberg-Rydberg nearest neighbour distribution yields an adventitious population of free electrons.  

These electrons, trapped by the NO$^+$ space charge drive an electron-impact avalanche.  The illuminated volume evolves on a 100 ns timescale to form a quasi-equilibrium of NO$^+$ ions, electrons and high-Rydberg molecules \cite{Haenel2017}.  Ensemble dynamics redistribute energy released in the avalanche to quench the electron temperature as well as the relative motion of ions and Rydberg molecules \cite{Sous2018,Sous2019}.

\subsection{Selective field ionization as a gauge of short-time dynamics }

We measure the distribution of electron binding energies in the Rydberg gas as a function of delay after $\omega_2$ by means of selected field ionization (SFI) spectroscopy \cite{gallagher2005rydberg}.  A Bertran high voltage power supply  biases a Behlke high voltage push-pull switch to generate a 3kV high voltage pulse. A 5 k$\Omega$ resistor in series with G$_1$ forms an RC circuit that generates a ramped electric field between G$_1$ and G$_2$ that rises at $\sim 0.8 $ V cm$^{-1}$ns$^{-1}$.  A polynomial function fit to the leading edge of the ramp voltage as a function of time transforms the detected electron signal waveform to a quantity that varies with the electric field between G$_1$ and G$_2$.  

An electric field with this rise time drives the superposition of $m_\ell$ states formed by the initial preparation of a particular $n_0f(2)$ Rydberg level to evolve diabatically through the Stark manifold, crossing a saddle point to form ions and free electrons when the field $F$ rises to a threshold amplitude, ${1}/{9n^4}$ in atomic units.  This separation of charge appears as an electron signal resonance that begins for $F =  (E_n(2)/4.59)^2$ V cm$^{-1}$, where $E_{n_0}(2)$ in cm$^{-1}$ expresses the binding energy of the Rydberg state with respect to a nitric oxide cation in rotational state, $N^+=2$.

As this $N^+=2$ wavepacket propagates, it traverses many intersections with states of matching total angular momentum and electronic $\times$ rotational parity built on the ground rotational state of NO$^+$.  These crossings confer sufficient $N^+=0$ character for amplitude to develop in a final state of free electrons and rotational ground state NO$^+$ ions, which can appear as a resonance at lower voltage, when the rising field passes an amplitude of $(E_n(0)/4.59)^2$ V/cm.

For an initial density of $10^{11}$ cm$^{-3}$, a field ionization ramp begun 300 ns after the $\omega_2$ samples the system in a state composed mainly of $n_0 f(2)$ Rydberg molecules as initially prepared, together with Rydberg molecules in other levels, $n$ and free electrons bound to the space charge of NO$^+$ ions formed early in the avalanche.

SFI ramps initiated with longer time delays sample distributions of $n_0$ Rydberg molecules altered by numerous collisions with avalanche electrons.  At a rate faster than that of $n$-changing interactions, these collisions mix $\ell$, populating degenerate manifolds of states at each $n$, $\ket{N^+=2} \ket{n,\ell}$.  These states of higher orbital angular momentum field ionize to produce electron waveforms reflecting the formation of ions in rotational states, $N^+=0$ and 2 at slightly higher field amplitudes. 

\subsection{Field-free propagation as a long-time gauge of plasma evolution }

Holding the electrostatic field to zero, we allow the photoexcited gas volume to travel with the propagation of the molecular beam to transit the far grid, G$_2$.  A bellows actuator longitudinally translates a carriage holding perpendicularly mounted grids G$_2$ and G$_3$ together with a microchannel plate detector assembly (Figure~\ref{fig:diagram}a).  Adjusting the flight distance from 3 to 60 mm allows field-free evolution times from 2 to 40 $\mu$s.  

As the plasma volume transits G$_2$, it encounters a static field set to a range from 30 to 200 V cm$^{-1}$.  This field extracts an electron signal that traces the width of the plasma in the in the direction of propagation.  As in the SFI experiment, we integrate this late-peak electron signal waveform to measure the total plasma and high-Rydberg electrons present for a given evolution time.  

We vary the  $\omega_1- \omega_2$ delay to regulate the  initial density of the Rydberg gas.  For a controlled density, we detect the change in the integrated late-peak signal in the presence of resonant mm-wave radiation to gauge the effect of early time Rydberg-Rydberg transitions on the fraction of the initial Rydberg gas that  evolves to a long-time ultracold plasma state of arrested relaxation.  

\subsection{UV-UV double-resonance and UV-UV-mm-wave triple-resonance spectroscopy }

A Virginia Diodes 10-400 GHz solid-state multi-band transmitter (VDI-Tx-S129) produces a tunable CW mm-wave field, $\omega_{\rm mm}$, with an energy density as high as 100 $\mu$W cm$^{-2}$.  This output originates from an on-board computer controlled Micro Lambda MLSE-1006 yttrium iron garnet (YIG) oscillator based frequency synthesizer, set for present purposes to a precisely determined primary frequency in the range from 10 to 20 GHz. Stages of frequency doubling (WR 5.1$\times$2R2) and tripling (WR 9.3$\times$3) produce continuously tunable output from 60 to 120 GHz. The multi-band transmitter broadcasts this output through a VDI-WR-10 35.5$\times$16.3 mm conical horn.   An $f/5$ teflon lens collects a portion of this mm-wave electromagnetic field and focuses it on the molecular beam with a maximum estimated power density of 25 $\mu$W cm$^{-2}$.  We regulate the output by means of cardboard attenuators placed between the horn and lens, and measure the power incident on the 2.5 cm diameter $\times$ 1 mm thick silica window of the vacuum chamber using a TK Instruments THz absolute power meter.  

For $\omega_1$ fixed on a transition to the A $^2\Sigma^{+} $ $N^{'}=0$ state, a scan of $\omega_2$ from 327 nm to 330 nm yields a double-resonant spectrum of $nf(2)$ Rydberg molecules that appears both in the SFI and late peak ultracold plasma signals.  Such scans in the presence of a CW mm-wave field forms a strong triple-resonance signal when the $\omega_2$ scan samples an $n_0f(2)$ state for which an appropriate $\Delta n = 1$ Rydberg-Rydberg resonance exists at $\omega_{\rm mm}$ (see below).   

Tuning $\omega_1$ and $\omega_2$ to a particular $n_0f(2)$ Rydberg state and scaning the mm-wave frequency, we can detect similar spectra both in the SFI and late peak ultracold plasma signals when the three radiation fields satisfy the triple-resonance condition.  We have sub-kHz control of the mm-wave frequency, so the observed variation of the triple-resonance signal with $\omega_{\rm mm}$ provides a means to assess the lineshapes of Rydberg-Rydberg transitions in the plasma.  

For some experiments we modulate the output power by applying 5 to 0 V half-wave voltage pulse to the TTL modulation input of the VDI-Tx-S129 transmitter.  A 5 V DC control voltage sets the transmitter output to zero.  Switching this input to 0 V sets the transmitter to full power.  We use this feature to delay the application of the mm-wave field with respect to $\omega_2$.

%%%%%%%%%%%%%%%%%%%%%%%%%%%%%%%%%%%%%%%%%%%%%%%%%%%%%%%%%%%%%%%%%%%%%%%%%%%%%%%%%%%%%%%%%%%%%%%%%%%%%%%
%%%%%%%%%%%%%%%%%%%%%%%%%%%%%%%%%%%%%%%%%%%%%%%%%%%%%%%%%%%%%%%%%%%%%%%%%%%%%%%%%%%%%%%%%%%%%%%%%%%%%%%
\section{Results}
\subsection{The effect of mm-wave radiation on the $\omega_2$ spectrum of ultracold plasma NO Rydberg series}

\subsubsection{UV-UV double-resonance }

Figure 1 shows double-resonant spectra obtained in transitions originating from the A $^2\Sigma^+$ $N^+=0$ state for $\omega_2$ wavelengths scanned from 328 to 329 nm in the absence of mm-wave radiation. Here, to obtain the upper spectrum we integrate the SFI signal formed by Rydberg molecules that retain the initial principal quantum number of the Rydberg gas, $n_0$. The lower spectrum represents the late peak signal formed as the illuminated volume transits $G_2$ after 20 $\mu$s of field-free flight. Both spectra show resonances in the $nf(2)$ series for principal quantum numbers, $n_0$, that extend from 36 to 49, converging to the NO$^+$ rotational level $N^+=2$ with orbital angular momentum $\ell=3$.  

%%%%%%%%%% w2 scan
\begin{figure}[h!]
        \includegraphics[width=.3\textwidth]{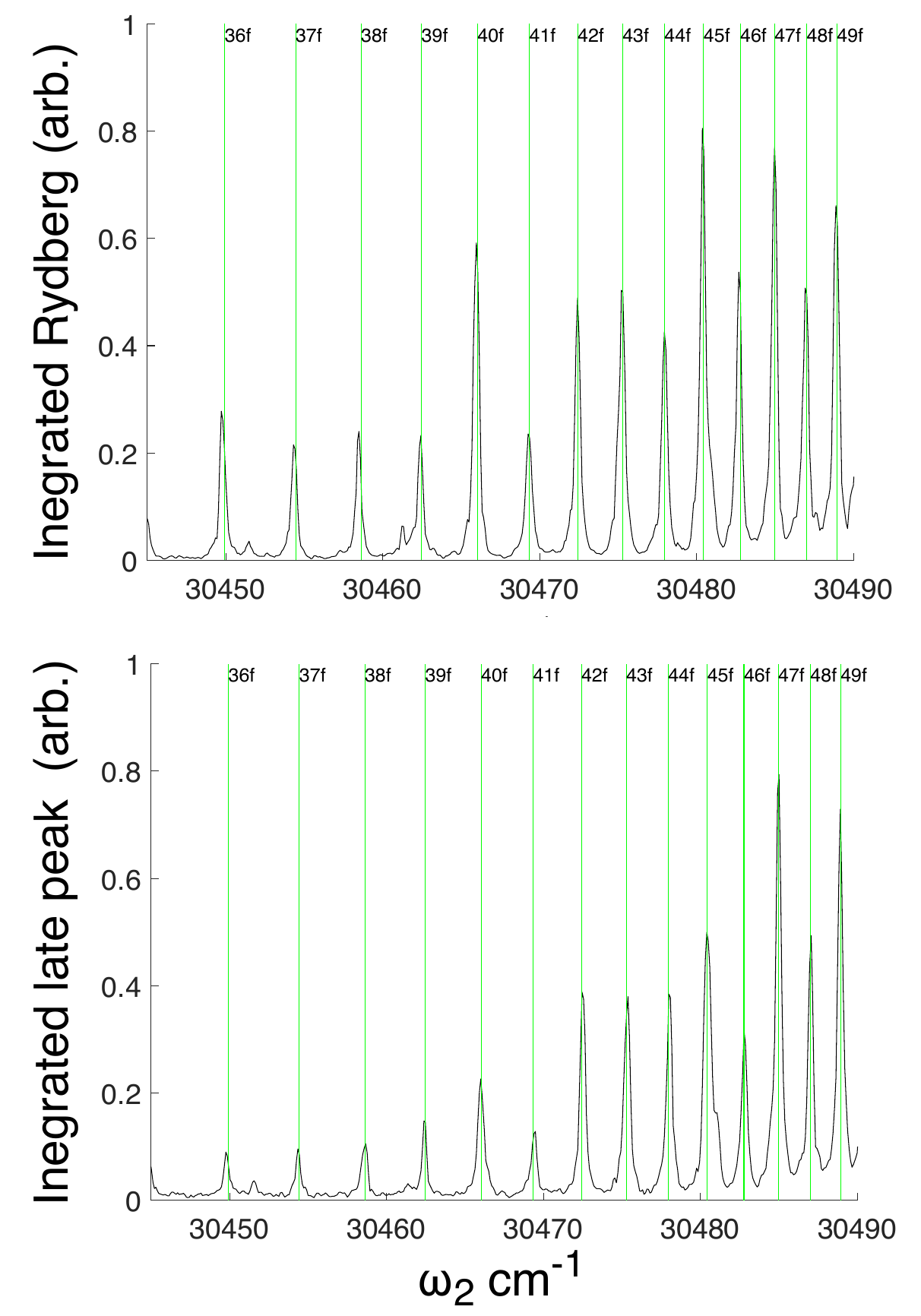} 
\caption{Field free Double-resonant $\omega_2$spectra of nitric oxide $n_0f(2)$ Rydberg states for $n_0$ from 38 to 49. (top) Electron signal integrated over Rydberg resonances in the SFI spectra. (bottom) Electron signal integrated over late peak after a flight time of 20 \us}\label{fig:no_mm}
\end{figure}
     
Note that, collecting either the SFI or late-peak electron signals, scans over this principal quantum number range yield Rydberg spectra that have comparable variations in intensity.  In particular, resonances in the upper half of these spectra alternate in intensity enhancing structure for $n_0= 40, ~45, ~47$ and 49, while suppressing $n_0 =41, ~46$ and 48.  Interloping structure appears between resonances for $n_0=36$ and 37 and $n_0=38$ and 39, seen easily in the SFI-detected spectrum, and detectable as well in the late-peak spectrum. 

\subsubsection{UV-UV-mm-wave triple resonance}

The application of a CW mm-wave field at particular frequencies from 70 to 110 GHz dramatically affects the intensities of certain particular features in the action spectra of $n_0f(2)$ Rydberg resonances for $n_0$ from 36 to 49, observed both by integrating the SFI signal and the late peak.  For each selected mm-wave frequency we usually see the enhancement of one particular principal quantum number in the $n_0f(2)$ spectrum.  In many cases, we also note a neighbouring resonance with diminished intensity.  Generally speaking, we find that enhancement occurs whenever the mm-wave frequency falls near the energy of a unit change in principal quantum number for a given $n_0f(2)$ resonance.  

\begin{figure}  
    \includegraphics[width=0.47 \textwidth]{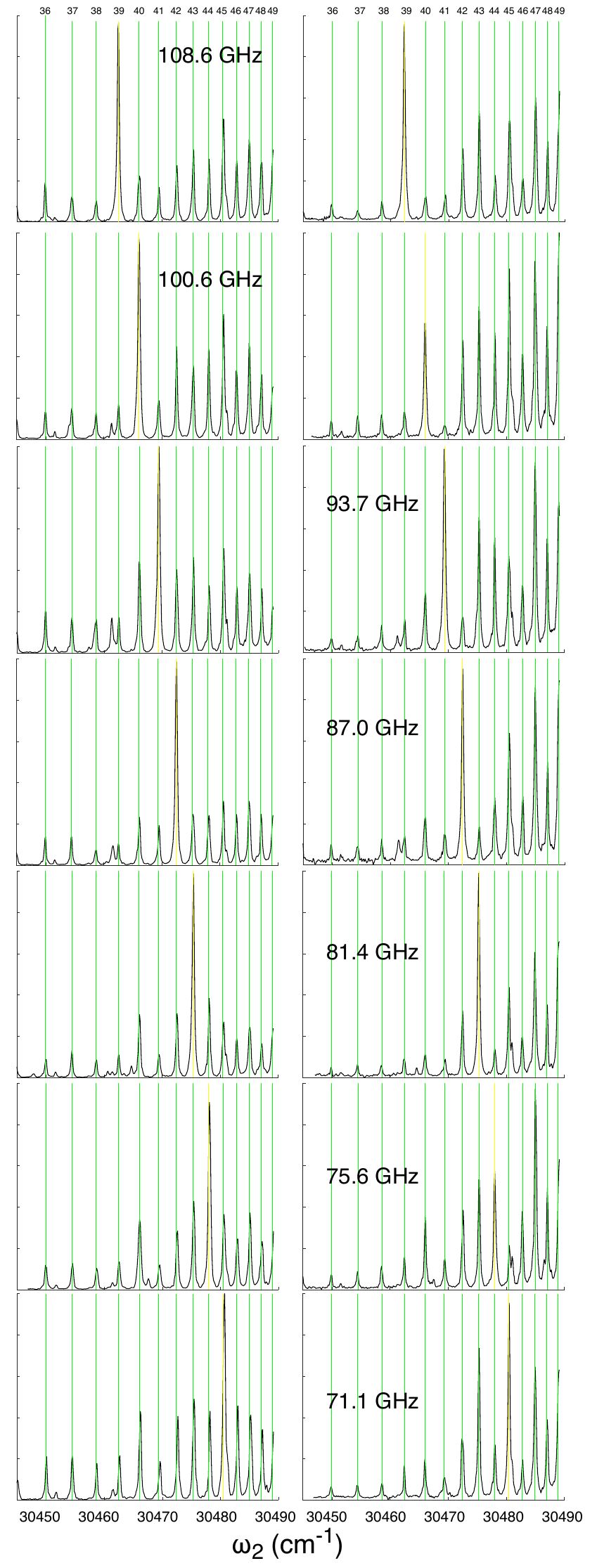}
   \caption{$\omega_2$ spectra as in Figure \ref{fig:no_mm}, with SFI (left) and late-peak (right) detection in mm-wave fields with the frequencies indicated. } 
   \label{fig:w2}
\end{figure}

\paragraph{108.6GHz  }
 Figure \ref{fig:w2} shows SFI and late-peak $\omega_2$ spectra as a function of $\omega_2$ scanned in the presence of a 100 $\mu$W mm-wave field with a frequency of 108.6 GHz.  Compare these spectra with those in Figure \ref{fig:no_mm}, recorded in the absence of a mm-wave field.  There, under field-free conditions, the $40f(2)$ resonance dominates the series of low-$n_0$ resonances.  The addition of a 108.6 GHz field enhances the intensity of the $39f(2)$ resonance by a factor of five in the SFI action spectrum and by a factor of ten in the late peak action spectrum compared with the field-free results.  The same 108.6 GHz field suppresses the $40f(2)$ feature by about a factor of 1.5 in the SFI spectrum and a factor of two in the late peak spectrum.

\paragraph{100.6 GHz }

A mm-wave field with a frequency of 100.6 GHz causes a similar pattern of enhancement and suppression in the SFI and late-peak $\omega_2$ spectra.  Note that a field of this frequency enhances the relative intensity of the $40f(2)$ resonance by factors of approximately two in both the SFI and late peak detected spectra. At the same time, it reduces the relative intensity of the $41f(2)$ resonance slightly in the SFI spectrum, and decreases it to nearly zero in the late-peak spectrum.

\paragraph{93.7 GHz }

A mm-wave field with a  frequency of 93.7 GHz increases the relative intensity of the $41f(2)$ resonance increases by a factor of five in the SFI spectrum and a factor of eight in the late peak spectrum. This mm-wave field suppresses the $42f(2)$ feature by two thirds and the $45f(2)$ resonance by half in the late peak spectrum, but causes no obvious suppression in the $\omega_2$ spectrum recorded by monitoring the SFI signal.  Note as well that this field enhances an interloping resonance that appears at 30461.46 cm$^{-1}$  in the SFI spectrum.  This resonance appears weakly as a new feature in the late peak spectrum in the presence of a 93.7 GHz field.

\paragraph{87.0 GHz }

A mm-wave field with a frequency of 87.0 GHz increases the relative intensity of the $42f(2)$ resonance by a factor of five in the SFI spectrum and by a factor of three in the late peak spectrum. We find the $43f(2)$ feature to be suppressed by half in the late-peak spectrum, but not suppressed at all in the SFI spectrum. Note that a 87.0 GHz mm-wave field again enhances the interloping resonance that occurs at 30461.65 $cm^{-1}$ in the SFI-detected $\omega_2$ spectrum.  It also causes this feature to appear again as a new resonance in the late peak spectrum.
 
\paragraph{81.4 GHz }
In the presence of a mm-wave field with a  frequency of 81.4 GHz, the relative intensity of $43f(2)$ resonance increases by a factor of two in the SFI-detected $\omega_2$ spectrum and increases by a factor of three in the late-peak $\omega_2$ spectrum. This field suppresses the $44f(2)$ feature by two thirds in the late peak spectrum, but not at all in the SFI-detected $\omega_2$ spectrum.  Note that the presence of an 81.4 GHz field causes a weak new feature to appear at 30464.71 cm$^{-1}$ between $39f(2)$ and $40f(2)$ in the SFI and late-peak-detected spectra.  This feature is absent in both classes of spectra under mm-wave field-free conditions.   
 
\paragraph{75.6 GHz }

A mm-wave field with a frequency of 75.6 GHz increases the intensity of the $44f(2)$ resonance by a factor of two in the SFI-detected $\omega_2$ spectrum but only slightly increases the intensity of this feature in the late-peak spectrum.  This mm-wave field suppresses the $45f(2)$ feature by two thirds in the late peak spectrum and one third in the SFI-detected $\omega_2$ spectrum. Note the creation of a weak new feature at $30467.56 cm^{-1}$ between $40f(2)$ and $41f(2)$ in both SFI and late peak spectra.  Again, this resonance does not appear in the field-free spectra. 
 
\paragraph{71.1GHz}

In the presence of a mm-wave field with a  frequency of 71.1 GHz, the relative intensity of the SFI-detected $45f(2)$ resonance grows by a factor of two.  This same resonance increases by a factor of three in the late-peak-detected $\omega_2$ spectrum.  Note that this field also increases the intensity of the $43f(2)$ resonance by a factor of two in the late-peak spectrum.  The 71.1 GHz field suppresses the $46f(2)$ resonance by a factor of three in the late-peak spectrum and slightly decreases the intensity of the $47f(2)$ resonance in the SFI-detected $\omega_2$ spectrum

\paragraph{Brief summary}

The $\omega_2$ spectra detected by integrating both the SFI and late-peak signals show a pattern of increased intensity in successive $n_0(2)$ resonances with decreasing mm-wave frequency.  In addition to these systematically resonant enhancements, we find that mm-wave fields of certain frequencies cause new, interloping features to appear at 30461.46, 30464.71 and 30467.56 cm$^{-1}$.  

Generally speaking, a field that enhances the strength of a resonance in the late-peak-detected $\omega_2$ spectrum for a principal quantum number, $n_0$, significantly depresses the intensity of the late-peak feature at $n_0 +1$.  We notice a similar but much weaker pattern of suppressed intensity in the $\omega_2$ spectra detected by monitoring the integrated SFI signal.

%%%%%%%%%%%%%%%%%%%%%%%%%%%%%%%% mmWave scan %%%%%%%%%%%%%%%%%%%%%%%%%%%%%%%%%%%%%%%%%%%%%%%%%%%%%%%%%%%
\subsection{Fixed $\omega_2$:  mm-wave excitation spectra of state-selected ultracold plasmas}

\begin{figure}  
    \includegraphics[width=0.421 \textwidth]{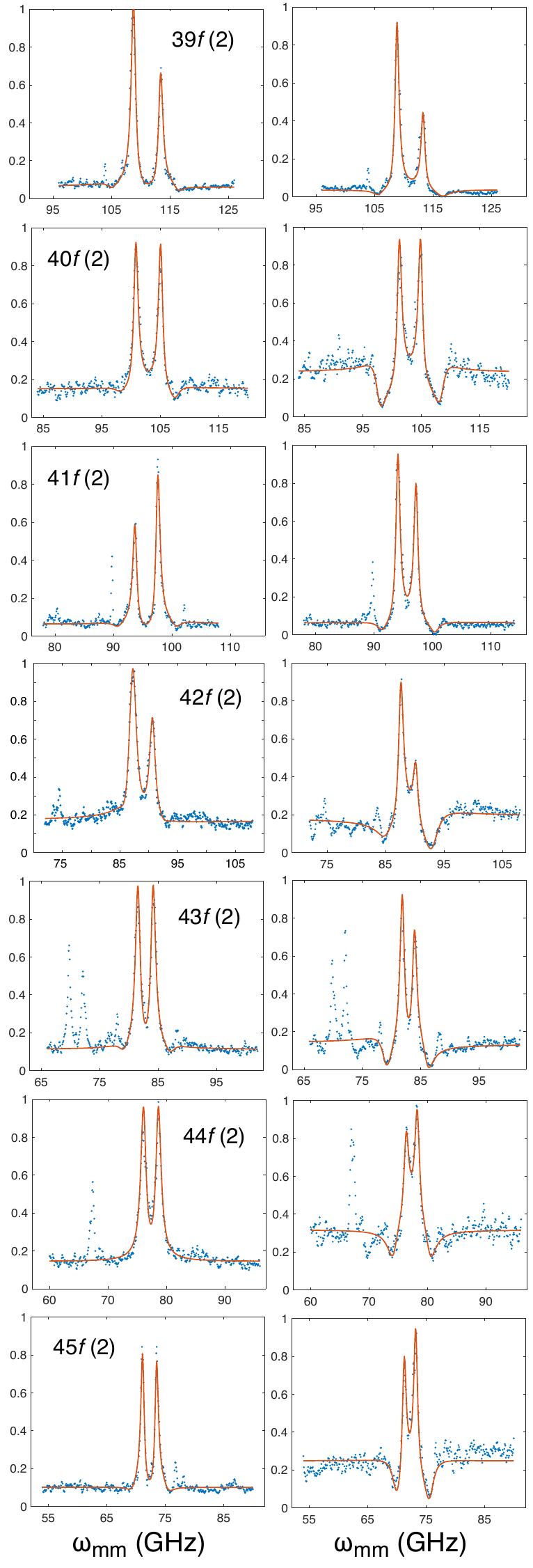}
 \caption{mm-wave action spectra measured by the SFI yield (left) and late-peak signal (right) for ultracold plasmas that evolve from Rydberg gases with $n_0$ from 39 to 45.}    
\label{fig:w3}
\end{figure}

For fixed $\omega_2$, tuned to a particular $n_0f(2)$ resonance, a scanned mm-wave frequency, $\omega_{\rm mm}$, produces a response in the electron signal, monitored either by integrating the SFI waveform collected 300 ns after $\omega_2$ or measuring the late peak after 20 $mu$s.  Figure \ref{fig:w3} shows mm-wave action spectra observed in the SFI signal (left) and late-peak signal (right) for ultracold plasmas that evolve from state-selected Rydberg gases with $n_0$ from 39 to 45.   

\paragraph{Ultracold plasma that evolves from $39f(2)$ } The top row of Figure \ref{fig:w3} shows the mm-wave excitation spectrum of the ultracold plasma that evolves from the $39f(2)$ Rydberg gas.  Here we see a distinctive pair of resonant features at 109 and 113 GHz with full-width at half-maximum (fwhm) of approximately 1 GHz.  These appear in almost identical fashion when detected in the SFI signal 300 ns after $\omega_2$ (left) and after 20 $\mu$s in the late peak (right).  Note as well a small, sharper satellite resonance at 104GHz in both spectra. The mm-wave spectrum gauged in the late peak also exhibits broader dips at 106GHz and 115 GHz. We see little if any sign of these dips in the spectrum observed after 300 ns using SFI detection. 
%At these two frequencies, small dips are notaceable in the SFI spectrum.

\paragraph{Ultracold plasma that evolves from $40f(2)$}  
The ultracold plasma that evolves from the $40f(2)$ Rydberg gas exhibits a closer pair of mm-wave resonances of similar fwhm at lower frequencies of 101 and 105 GHz.  Over this scan of 40 GHz, we find no satellite peaks.  But the late-peak mm-wave spectrum forms broad dips with centre frequencies of 98 and 107 GHz.  A hint of the higher-frequency dip appear in the short-time SFI-detected spectrum.   

\paragraph{Ultracold plasma that evolves from $41f(2)$} 
The same pattern of 2 GHz fwhm doublets appears with narrower spacing at lower frequencies of 94 and 97 GHz in transitions from $\omega_2$-selected $41f(2)$.  Here we also see a small satellite peak at 89 GHz in both SFI and late-peak spectra.  Broader dips occur at 91 GHz and 100 GHz occur prominently in the late-peak spectrum.  The SFI-detected spectrum shows a suggestion of this structure.   

\paragraph{Ultracold plasma that evolves from $42f(2)$} 
The mm-wave spectra from $42f(2)$ again feature prominent doublets, now at 87 GHz and 90 GHz.  Here we see a weak additional peak at 74 GHz in both spectra.  Broad dips appear at 85 GHz and 92 GHz in the late-peak signal detected after a flight time of 20 $\mu$s.  

\paragraph{Ultracold plasma that evolves from $43f(2)$} 
The plasma prepared from a $43f(2)$ Rydberg gas yields a similar doublet of mm-wave excitation features at 82 GHz and 84 GHz in both SFI and late spectra.  The spectrum collected monitoring the late-peak signal also shows broader flanking dips at 80 GHz and 86 GHz.  Both the SFI and late-peak spectra show a small satellite peak at 78 GHz and a second pair of resonances at 70 GHz to 73 GHz, flanked by noticeable dips in the late-peak excitation spectrum.

\paragraph{Ultracold plasma that evolves from $44f(2)$} 
The mm-wave excitation spectra observed for a plasma that evolves from the $44f(2)$ Rydberg gas exhibits a familiar doublet, moved to lower frequencies of 76 GHz and 78 GHz.  Here, the spectrum sampled in the late-peak signal shows narrower and weaker dips at 74 GHz and 81 GHz.  Both spectra contain an isolated peak at 67 GHz, accompanied by a dip at 70 GHz in the late-peak spectrum.  

\paragraph{Ultracold plasma that evolves from $45f(2)$} 
We find the lowest-frequency, narrowest-spaced doublet of mm-wave transitions in this series in the plasma that originates from a Rydberg gas selected to originate from $45f(2)$.  Excitation spectra collected at short time in the SFI signal and after 20 $\mu$s in the late peak exhibit resonances at at 71 GHz and 73 GHz.  Pronounced dips at 70 GHz and 75 GHz flank these features in the late-peak mm-wave spectrum.  Interestingly, notice that the frequency positions of the prominent doublet here closely matches the satellite structure observed in the plasma that evolves from the $43f(2)$ Rydberg gas.

\paragraph{Summary}

Ultracold plasmas evolving from $n_0f(2)$ Rydberg gases for $n_0$ from 39 to 45 show a consistent pattern of positive going doublets, detected both in the SFI signal and the integrated late peak.  Both the transition energies and the spacing of these doublets decrease systematically with increasing $n_0$.  

Broad dips flank these doublets in the mm-wave spectrum measured by the late-peak signal.  The separation of these features also falls uniformly with increasing $n_0$.  Often the position of a positive-going resonance for a given value of $n_0$ occurs at the frequency of the upper dip in the excitation spectrum of the plasma that evolves from $n_0+1$.  These dip resonances are largely absent in $\omega_{\rm mm}$ excitation spectra recording the SFI spectra.  

This appears to explain the suppression of $(n_0+1)f(2)$ features in late-peak $\omega_2$ spectra recorded in the presence of mm-wave fields chosen to enhance $n_0f(2)$ resonances.  Consistent with the trend observed in these spectra, resonant dips prominent in the signal recorded after an evolution of 20 $\mu$s appear weakly if at all in short-time SFI-detected spectra.  

Nearly all of these mm-wave excitation spectra show additional positive-going satellite resonances.  In some cases these satellite resonances mirror one or both components of a principal doublet observed for a different $n_0$.

\subsection{Time-dependence of mm-wave excitation spectra}

Figures \ref{fig:w2} and \ref{fig:w3} display mm-wave frequency dependent changes in electron-signal intensities associated with UV excitation from A $^2\Sigma^+$ $N^+=0$ to Rydberg states with $n_0$ from 39 to 45, as well as mm-wave excitation from selected $n_0f(2)$ states.  These changes occur owing to the application of a CW mm-wave field during a period of 300 ns between $\omega_2$ excitation and the SFI ramp, and to nearly an equivalent extent in the presence of the same field for the 20 $\mu$s between $\omega_2$ and passage through G$_2$ to form the late peak.  

The similarity of excitation spectra whether monitoring SFI signal or the late peak raises the question of when during these intervals of time does interaction of the mm-wave field with the ultracold plasma act to produce the observed $\omega_{\rm mm}$ resonant effects.  

To explore this question, we have used the amplitude modulation feature of our multi-band transmitter to delay the application of the mm-wave field by various intervals of time after $\omega_2$ excitation to form the nitric oxide Rydberg gas.   

\begin{figure}[h!]
       \includegraphics[width=.37\textwidth]{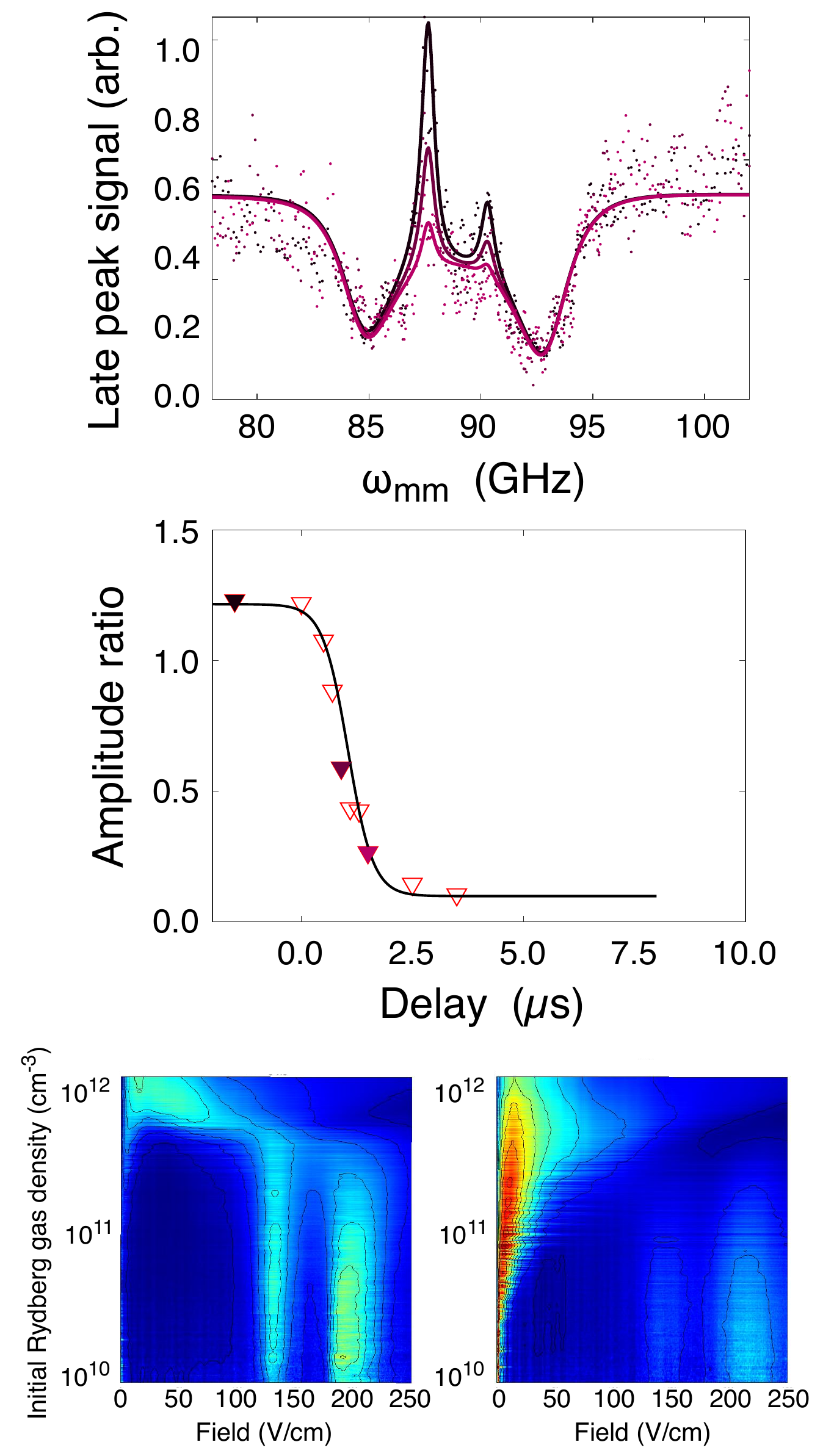}
     \caption{Time-dependent mm-wave spectra of 42f(2) state integrated over late-peak signal. (top frame) Spectra taken -1.8 $\mu$s, 0.8 $\mu$s and 1.4 $\mu$s relative to $\omega_2$ laser excitation respectively with color keyed to delay time in the (middle frame) ratio of signal integrated over the $42f(2) \to 43g(2)$ resonance peak,  to that integrated over the $42f(2) \to 43d(2)$ dip in the mm-wave spectra as a function of mm-wave delay time.  (bottom frame) Four thousand selective field ionization spectra arranged by decreasing initial Rydberg gas density, showing the electron signal as a function of electrostatic field, ramped from 0 to 250 V cm${-1}$ with ramp delays of 0 and 450 ns.  Signal observed for high densities at low voltage represents weakly bound high-Rydberg molecules and electrons bound to the NO$^+$ space charge.  The initial Rydberg molecules, in this case in the $44f(2)$ state, form the distinctive pair of bands at 125 and 180 V cm$^{-1}$, the field-ionization thresholds to produce NO$^+$ in rotational states $N^+= 0$ and 2, respectively.  Notice the shift to higher field and decrease in intensity as electron-Rydberg collisions mix $\ell$.  The $n_0 \ell(2)$ signal decreases further owing to early inelastic scattering and predissociation to form just a faint trace after 2 $\mu$s.   }
  \label{fig:mm_delay}   
   \end{figure}

The upper frame of Figure \ref{fig:mm_delay} shows a set of mm-wave scans over the resonant doublet and flanking dips observed for the plasma that evolves from a $42f(2)$ Rydberg gas, as pictured in the middle frame of Figure \ref{fig:w3} using three different delays, $\Delta t_{\rm mm}$, with respect to the $\omega_2$ excitation pulse,  -1.8 $\mu$s, 0.8 $\mu$s and 1.4 $\mu$s.  Here, we have scaled the data in each case to to match the amplitude of the evident dips.  

Recognizing the evident asymmetry of these features, we fit the data obtained for the mm-wave field started -1.8 $\mu$s before $\omega_2$ to a sum of four Fano profiles, two that describe the positive-going central features and two negative-going profiles that describe the flanking dips.  A set of 10 scans for $\omega_{\rm mm}$ delays varying from $\Delta t_{\rm mm}= -1.8 ~\mu$s to 3 $\mu$s yields best fitting amplitude ratios as plotted in the lower frame of Figure \ref{fig:mm_delay}.  The three curves in the upper frame of Figure \ref{fig:mm_delay} show the best fit simulations for the corresponding three delay times. 

Note that, scaled to a constant relative dip, the positive going amplitude observed in the late-peak $\omega_{\rm mm}$ spectrum decays to near zero on the timescale of the avalanche from Rydberg gas to plasma, about 1 $\mu$s.  For short variations in $\Delta t_{\rm mm}$, the amplitude of the dips is nearly constant, and this fall in the peak-to-dip ratio approximately describes the time decay in mm-wave enhancement of $n_0f(2)$ resonant features.  Over longer delays, we find that the absolute amplitudes of the depletion features diminish on 10 $\mu$s timescale in direct proportion to the length of the interval between the start of the $\omega_{\rm mm}$ square wave and the start of the late-peak signal. 

 \subsection{Power dependence of mm-wave excitation spectra}

 The Virginia Diodes solid-state multi-band transmitter forms a much narrower source of  pure CW mm-wave frequencies than the evident widths of $\omega_{\rm mm}$ Figures \ref{fig:w3} and \ref{fig:mm_delay}.  It thus remains to explain the persistent $\sim 1$ GHz, seemingly asymmetric lineshapes of these resonant features.   
 \begin{figure}[h!]
       \includegraphics[width=.32\textwidth]{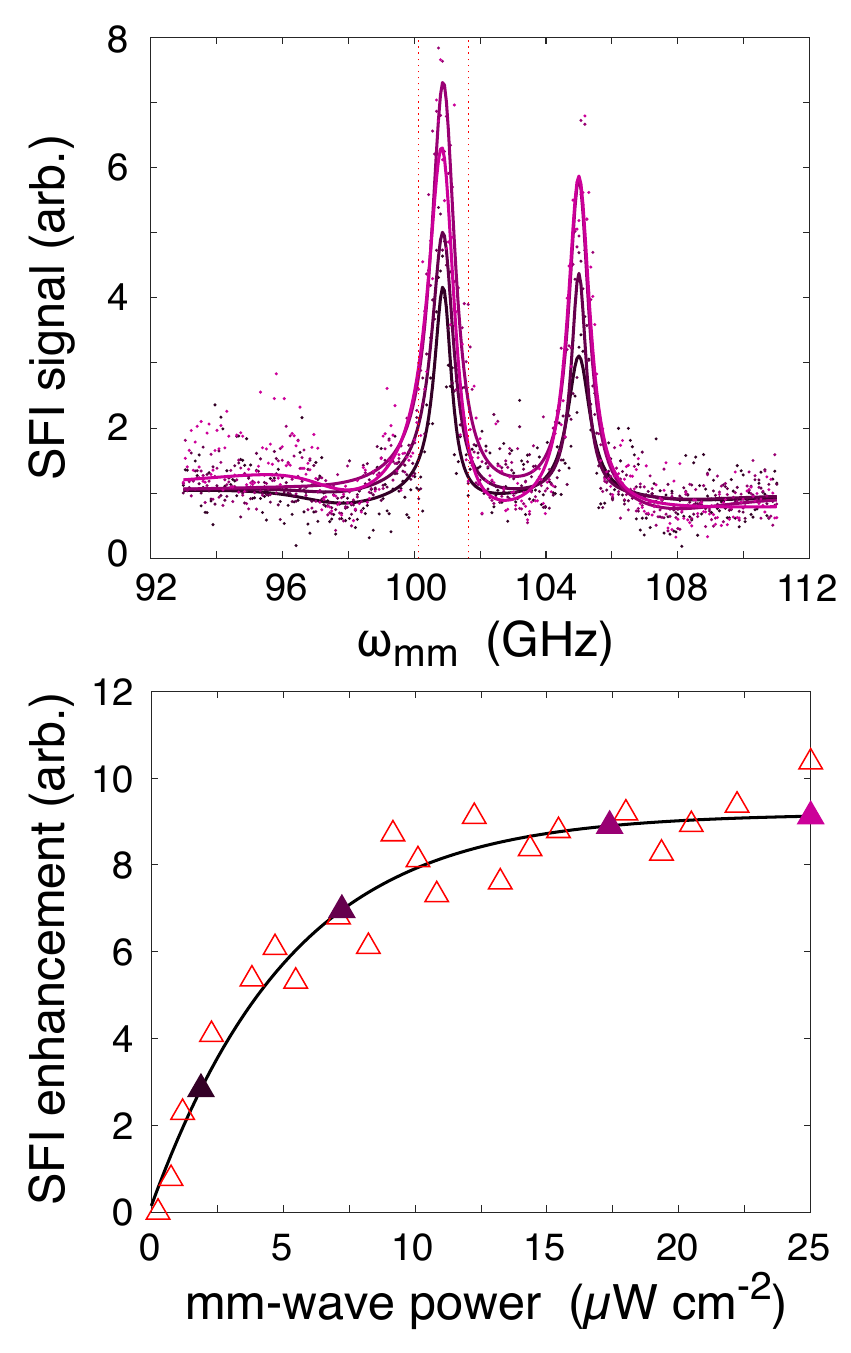}
     \caption{Power-dependence measurement of 40f(2) with SFI signal in  mm-wave spectra. (a) signal integrated over 100-102 GHz in with different mm-wave output power. (b)mm-wave spectra at 4 different mm-wave power indicated by the consistent colors of solid triangles in (a)} \label{fig:power}
   \end{figure}

Rydberg-Rydberg transitions exhibit enormous dipole transition matrix elements \cite{gallagher2005rydberg}, which presents the issue of power broadening.  The lower frame of Figure \ref{fig:power} integrates the resonant enhancement of the SFI signal observed for the plasma that forms from the $40f(2)$ Rydberg gas for $\omega_{\rm mm}$ from 100 to 102 GHz at incident powers from 0 to 25 $\mu$W cm$^{-2}$.  Over this range, the absorption rises from a regime of linear absorption to approach saturation.  The upper frame of Figure \ref{fig:power} shows $\mu$W cm$^{-2}$ spectra at incident mm-wave intensities marked by filled triangles.  Note that power broadening has no effect on the observed linewidths over this full range.

% can be a 10mW fluctuation of power with frequency. This is why one peak is saturated while the other one is not. 

%%%%%%%%%%%%%%%%%%%%%%%%%%%%%%%%%%%%%%% Discussion %%%%%%%%%%%%%%%%%%%%%%%%%%%%%%%%%%%%%%%%%%%%%%%%%%%%

\section{Discussion}
\subsection{Rydberg spectroscopy of the nitric oxide molecular ultracold plasma}

The formation of a durable ultracold plasma marks the energy positions of long-lived high Rydberg states of nitric oxide accessible in UV-UV double resonance from a state of zero angular momentum neglecting spin.  The addition of a mm-wave radiation field tunable from 60 to 120 GHz drives Rydberg-Rydberg transitions that enhance or diminish the electron signal associated with this plasma.  These results shed light on the high-Rydberg spectroscopy of NO.  They also point with quantum state resolution to dynamical factors that shape the evolution to an ultracold molecular plasma and determine its stability.  We begin with an analysis of the spectroscopy.   

\subsubsection{UV-UV double-resonance and the $n_0f(2)$ Rydberg spectrum of NO}

Angular momentum selection rules allow UV-UV double resonance electronic transitions via the A $^2\Sigma^+$ $N'=0$ intermediate state of nitric oxide to reach all Rydberg states for which, $N$, the total angular momentum neglecting spin equals 1.  This optical excitation process forms neutral Rydberg gases of varying stability with respect to predissociation to atoms:
\begin{equation}
{\rm NO} \ket{n_0,\ell} \rightarrow {\rm N(^4S )+ O(^3P)}
\end{equation}
Many previous studies have recognized that the rate of predissociation, which depends sensitively on Rydberg orbital angular momentum, $\ell$, limits transitions readily detected by ionization from A $^2\Sigma^+$ $N'=0$ to those in the $N=1$ $nf$ Rydberg series that converges to the NO$^+$ rotational level, $N^+=2$ \cite{Vrakking1995,Bixon,Remacle1998,Murgu2001,patel2007rotational}.

This hydrogenic series, $n_0f(2)$, defines a particular subset of nitric oxide Rydberg gas initial states that evolve to form a durable ultracold plasma.  The integrated amplitude of the selective field ionization (SFI) signal, sampled 300 ns after $\omega_2$ measures the resonant position of each resonance and the effectiveness by which the associated short-time dynamics of avalanche ionization forms a bound distribution of NO$^+$ - $e^-$ charge.  

On a far longer timescale, after 20, 40 or as much a 400 $\mu$s \cite{Schulz-Weiling2016,schulz2016evolution,Haenel2017}, an electron signal late peak forms when the illuminated volume transits a detection plane.  The amplitude of this signal gauges the effectiveness with which each $n_0f(2)$ Rydberg gas evolves to yield an ultracold plasma state of arrested dissipation.  As we can see from Figure \ref{fig:no_mm} the long-time, late-peak signal presents much the same intensity pattern as the SFI spectrum sampled after only 300 ns.  From this we can conclude that dynamical processes occur very early in the avalanche of the Rydberg gas to determine the charge density of the plasma sampled at a much later time.

\subsubsection{UV-UV-mm-wave triple-resonance and the assignment of $n_0f(2) \rightarrow (n_0 \pm 1)g(2)$ Rydberg - Rydberg transitions}

A CW mm-wave field, tuned to a succession of specific points in the range of frequency from 120 and 60 GHz, enhances the intensity of $n_0f(2)$ resonances in the $\omega_2$ spectrum for $n_0$ from 39 to 45.  The sequence of spectra in Figure \ref{fig:w2} defines the set of mm-wave resonant frequencies that decreases as a slightly sub-linear function of $n_0$ (cf. Figure \ref{fig:QD}).  This set of $\omega_{\rm mm}$ frequencies conforms almost exactly with the set of transition energies from each $n_0f(2)$ level to the neighbouring $(n_0+1)g(2)$ state calculated simply from Rydberg formula with constant quantum defects, $\delta_f=0.023$ and $\delta_g=0.003$ \cite{Murgu2001}.

\begin{table}[h!]
	\centering
	\caption{Assignment of $n_0f(2)$ positions and $n_0f(2)$ to $(n_0 \pm 1)g(2)$ Rydberg-Rydberg transitions from $\omega_2$ and $\omega_{\rm mm}$ spectra }
	\label{tab:f-g}
	\resizebox{0.49\textwidth}{!}{%
		\begin{tabular}{ccrrrr}
			\toprule
	\multirow{2}{*}{ $\omega_2 $ $cm^{-1}$	} & \multirow{2}{*}{$n_0f(2)$} &\multicolumn{2}{c}{$n_0f(2) \rightarrow (n_0+1)g(2)$ } & \multicolumn{2}{c}{$n_0f(2) \rightarrow (n_0-1)g(2)$} \\ 
		&                        & meas(GHz)        & calc(GHz)        & meas(GHz)      & calc(GHz)                \\ 
			\midrule
30462.46&	$39f(2)$&	108.84 & 109.04 & 113.22 & 113.14 \\
30466.03&	$40f(2)$&	101.18 & 101.16 & 104.91 & 104.76 \\
30469.35&	$41f(2)$&	94.08  & 94.01  & 97.25  & 97.19  \\
30472.42&	$42f(2)$&	87.61  & 87.53  & 90.22  & 90.33  \\
30475.29&	$43f(2)$&	81.86  & 81.63  & 83.92  & 84.10  \\
30477.96&	$44f(2)$&	76.33  & 76.25  & 78.26  & 78.43  \\
30480.46&	$45f(2)$&	71.25  & 71.33  & 73.23  & 73.25  \\       
			\bottomrule
		\end{tabular}%
	}
\end{table}

For a set of $\omega_2$ frequencies fixed on specific $n_0f(2)$ resonances, scans of $\omega_{\rm mm}$ shown in Figure \ref{fig:w3} confirm these assignments.  Here we see clear doublet positions that fit precisely with the energies calculated for transitions from $n_0f(2)$ to $(n_0+1)g(2)$ and $(n_0-1)g(2)$.  Combinations of energies for the mm-wave transitions $(n_0-1)f(2)$ to $(n_0)g(2)$ and $(n_0+1)f(2)$ to $(n_0)g(2)$ accurately reflect the binding energy differences between  $(n_0-1)f(2)$ and $(n_0+1)f(2)$ measured by resonant positions in the ultraviolet $\omega_2$ spectrum.

Table \ref{tab:f-g} summarizes the $\omega_2$ positions of the $n_0f(2)$ Rydberg states for $n_0$ from 39 to 45, and compares measured mm-wave $n_0f(2)$ to $(n_0 \pm 1)g(2)$ Rydberg-Rydberg transition energies with simple calculations assuming constant quantum defects.

\subsubsection{Dips in the $\omega_{\rm mm}$ spectrum mark Rydberg-Rydberg transitions from $n_0f(2)$ to dissociative $(n_0 \pm 1)d(2)$ states}

Distinctive dips flank all of the $n_0f(2) \rightarrow (n_0 \pm 1)g(2)$ Rydberg-Rydberg resonances in the late-peak excitation spectra plotted on the right in Figure \ref{fig:w3}.   The frequency positions of features fit well in each instance with transitions from the selected $n_0f(2)$ state to the adjacent $(n_0 \pm 1)$ levels of the $d\sigma$ series converging to NO$^+$, $N^+=2$ with a quantum defect, $\delta_d = -0.057$ \cite{Murgu2001}.  

 \begin{table}[h!]
	\centering
	\caption{Assignment of $n_0f(2)$ to $(n_0 \pm 1)d(2)$ Rydberg-Rydberg transitions from $\omega_{\rm mm}$ spectra }
	\label{tab:f-d}
	\resizebox{0.49\textwidth}{!}{%
		\begin{tabular}{ccrrrr}
			\toprule
			\multirow{2}{*}{ $\omega_2 $ $cm^{-1}$} & \multirow{2}{*}{$n_0f(2)$} &\multicolumn{2}{c}{$n_0f(2) \rightarrow (n_0+1)d(2)$ } & \multicolumn{2}{c}{$n_0f(2) \rightarrow (n_0-1)d(2)$} \\ 
			&                        & meas(GHz)        & calc(GHz)        & meas(GHz)      & calc(GHz)                \\ 
			\midrule
30462.46&	$39f(2)$&	116.42 & 115.19 & 105.78 & 105.96 \\
30466.03&	$40f(2)$&	108.10 & 106.87 & 98.13  & 98.12  \\
30469.35&	$41f(2)$&	100.12 & 99.33  & 91.32  & 91.03  \\
30472.42&	$42f(2)$&	93.07  & 92.49  & 85.11  & 84.61  \\
30475.29&	$43f(2)$&	86.10  & 86.25  & 79.04  & 78.78  \\
30477.96&	$44f(2)$&	80.34  & 80.57  & 74.24  & 73.47  \\
30480.46&	$45f(2)$&	75.65  & 75.37  & 69.42  & 68.63   \\              
			\bottomrule
		\end{tabular}%
	}
\end{table}

Note, however that mm-wave transitions originating from $n_0 \ell(2)$ states of any higher orbital angular momentum $\ell$, formed as a result of electron-Rydberg collisional $\ell$-mixing also fall within the 2 GHz bandwidth of these resonant dips.  Note as well that $\omega_{\rm mm}$ transitions to $(n_0 \pm 1) d(2)$ deplete the 20 $\mu$s late-peak ultracold plasma spectrum nearly to zero, while having a barely perceptible effect on the short-time signal of the avalanching plasma detected by SFI.

\begin{figure}[h!]
       \includegraphics[width=.46 \textwidth]{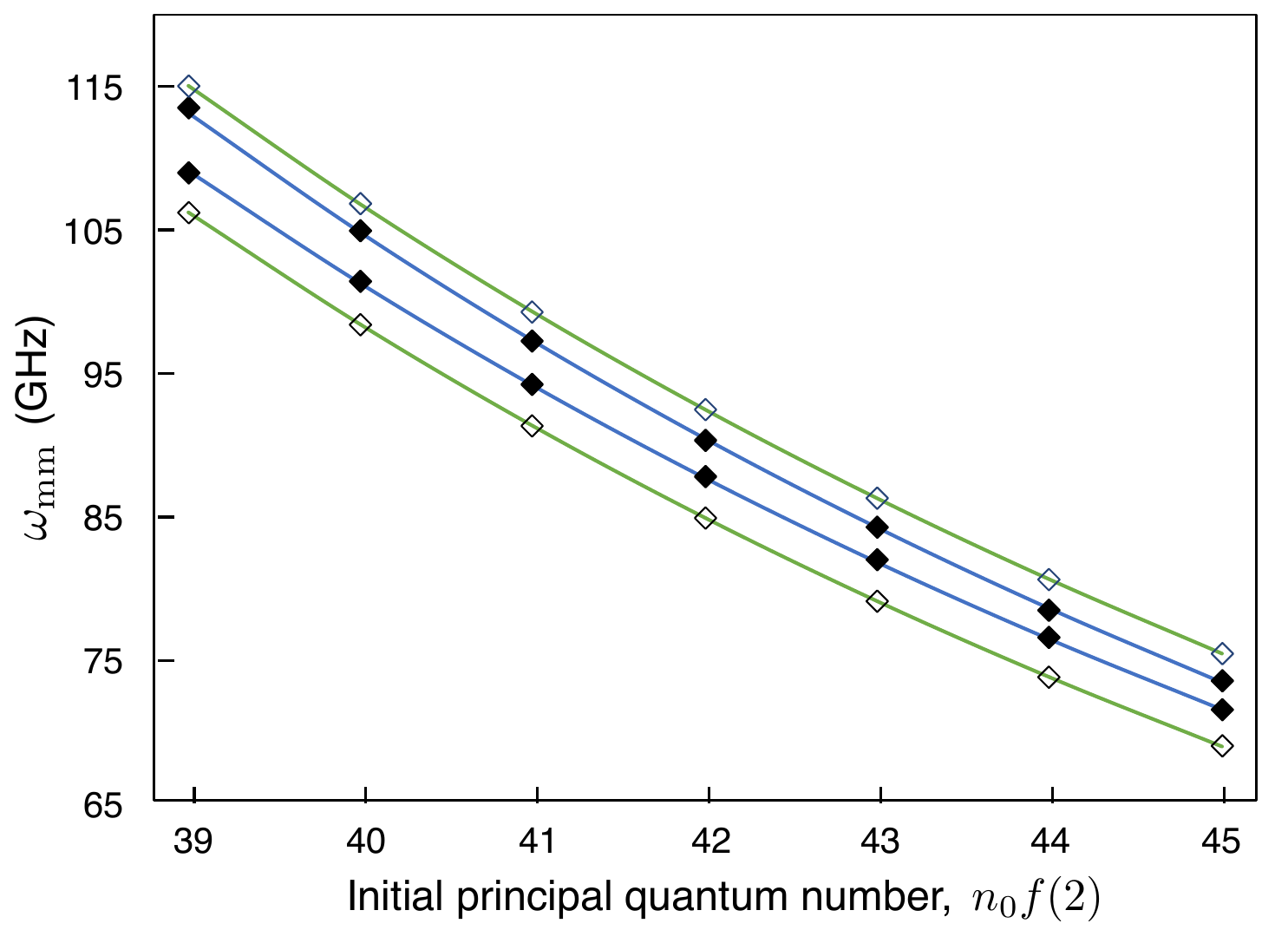}
     \caption{Measured positions of $\omega_{\rm mm}$ resonances for $n_0f(2) \rightarrow (n_0\pm 1)g(2)$ ($\blacklozenge$) and $n_0f(2) \rightarrow (n_0\pm 1)d(2)$ ($\lozenge$) transitions compared with lines connecting the calculated displacements of the successive sets of $g(2)$ and $d(2)$ states from the corresponding member of the $n_0f(2)$ series, for $n_0$ from 39 to 45. }   \label{fig:QD}
   \end{figure}

\begin{table*}
%	\centering
	\caption{Assignments of interloping $n_0\ell (N^+)$ positions for $\ell = 1, 2, 3$, and Rydberg-Rydberg transitions from $\omega_2$ and $\omega_{\rm mm}$ spectra.  Asterisks below refer to transitions for which calculated energies fall below the lowest scanned frequency.}	
	\label{tab:interloper}
%	\resizebox{.9\textwidth}{!}{%
		\begin{tabular}{cccccccc}
			\toprule
%\Xhline{2\arrayrulewidth} 		
	\vspace{2 pt}		\multirow{2}{*}{ $\omega_2 $ $cm^{-1}$} & {}\multirow{2}{*}{$n_0 \ell (N^+)$} &\multirow{2}{*}{$(n_0 +1)g (N^+)$}  &\multicolumn{2}{c}{$n_0 \ell (N^+) \rightarrow (n_0+1)g(2)$ } & \multirow{2}{*}{ $(n_0-1) g (N^+)$ }& \multicolumn{2}{c}{$n_0 \ell (N^+)  \rightarrow (n_0-1)g(2)$ }  \\ %\cline{3-10} 
			&       &                & meas(GHz)        & calc(GHz)  &         & meas(GHz)      & calc(GHz)             \\ 
%\Xhline{1\arrayrulewidth}		
			\midrule  
			\multicolumn{1}{c}{$\omega_2$ scan}   & \multicolumn{7}{c} {}  \\                                                                                                  
30447.98 & $39p(0)$ &&& & $38g(1)$ &81.40&83.28\\
30458.35 & $40d(1)$ & $41g(1)$ &93.70&93.66&&&\\
30461.46 & $43p(0)$ &&& & $42g(1)$ &93.70&92.56\\
30461.65 & $41d(1)$ & $42g(1)$ &87.00&87.05&&&\\
30464.26 & $44p(0)$ &&& & $43g(1)$ &93.70&94.37\\
30464.71 & $42d(1)$ & $43g(1)$ &81.40&81.05&&&\\
%30466.30 & $40d(2)$ &&& & $39g(2)$ &110.10&112.26\\ This is nd(2) interloper, we may not put in
30467.56 & $43d(1)$ & $44g(1)$ &75.60&75.59&&&\\
\multicolumn{2}{c} {$\omega_{\rm mm}$  scan: 30462.46,  $\left[39f(2)\right ]$} & \multicolumn{6}{c} {}  \\
30461.65 & $41d(1)$ &$42g(1)$&*& 87.05& $40g(1)$ &103.92&104.15\\
\multicolumn{2}{c}{$\omega_{\rm mm}$  scan: 30469.35,  $\left[41f(2)\right]$}& \multicolumn{6}{c} {}  \\
30467.56 & $43d(1)$ &$44g(1)$&*& 75.59& $42g(1)$ &89.79&90.13\\
\multicolumn{2}{c}{$\omega_{\rm mm}$  scan: 30475.29,  $\left[43f(2)\right]$} & \multicolumn{6}{c} {}  \\
30472.53 & $45f(1)$ & $46g(1)$ &69.87&71.34 & $44g(1)$ &72.12&73.24\\
30472.70 & $45d(1)$ & $46g(1)$ &66.81&66.06 & $44g(1)$ &78.15&78.52\\
\multicolumn{2}{c}{$\omega_{\rm mm}$  scan: 30477.96,  $\left[44f(2)\right]$} & \multicolumn{6}{c} {}  \\
30477.20 & $47d(1)$ &$48g(1)$&*& 58.03& $46g(1)$ &66.93&68.82\\
%\Xhline{1.5\arrayrulewidth}		
			\bottomrule
					%*{\footnotesize{Below lowest scanned frequency}}
		\end{tabular} 
%	}
\end{table*}

Table \ref{tab:f-d} lists the mm-wave $n_0f(2)$ to $(n_0 \pm 1)d(2)$ Rydberg-Rydberg transition frequencies measured for these bands compared with positions estimated from the Rydberg formula assuming constant quantum defects.  Note the reversal:  For this range of $n_0$, the calculated energy of the transition from $n_0f(2)$ to the level, $(n_0+1)d$, exceeds that for the transition to $(n_0-1)d$

In summary, we assign the very evident pairs of doublets that appear in these $\omega_{\rm mm}$ scans of resonances in the ultracold plasma response to transitions from $n_0f(2)$ states to adjacent Rydberg levels, $(n_0\pm 1)g(2)$ and $(n_0\pm 1)d(2)$ for $n_0$ from 39 to 45.  Figure \ref{fig:QD} confirms this assignment by showing the measured position of each resonance compared to a plot of the calculated displacement of each $(n_0\pm 1)g(2)$ and $(n_0\pm 1)d(2)$ state from the corresponding selected $n_0f(2)$ state.

\subsubsection{Additional resonances evident in the $\omega_{\rm mm}$ spectra and isolated regions of the $\omega_2$ spectra}

The $\omega_2$ spectra in Figure \ref{fig:w2} show evidence of additional structure between $n_0=39$ and 40, and in the shoulder of the resonance for $n_0=44$  that varies in the presence of mm-wave fields of particular frequency.  Similarly, the $\omega_{\rm mm}$ spectra in Figure \ref{fig:w3} exhibit numerous instances of additional resonances for fixed values of $\omega_2$ tuned to $n_0f(2)$ states with $n_0=39$, 41, 43 and 44.

As made evident by Figure \ref{fig:QD} together with Tables \ref{tab:f-g} and \ref{tab:f-d}, a simple Rydberg model describes the positions of states with $n_0$ in the range from 39 to 45 quite well in terms of separable rotational quantum number and constant quantum defect for $d$, $f$, and $g$ series.  Considering series that converge to $N^+=0$ and 1 as well as 2, we can identify intervals of $\omega_1 + \omega_2$ energy at which members of these interloping series coincide with $n_0f(2)$ states.  This occurs particularly in stroboscopic regions where Rydberg intervals approach rotational energy spacings.  

Gauging the calculated energy separation between candidate interloping $n_0p(N^+)$ and $n_0d(N^+)$ and adjacent $(n_0 \pm 1)g(N^+)$ states, we can account for all the other resonant features that appear in Figures \ref{fig:w2} and \ref{fig:w3}.  Table \ref{tab:interloper} assigns all of the transitions observed owing to interloping $p$, $d$ and $f$ states with total angular momentum neglecting spin of $N=1$.  It seems likely that all levels with $\ell<3$ borrow lifetime from neighbouring $n_0f(2)$ states, and like those principal $\omega_2$ resonances, add to the plasma signal detected in the SFI spectrum or the late peak when promoted to adjacent $(n_0 \pm 1)g$ states.

 \subsection{Intensity of mm-wave resonances as a probe of molecular ultracold plasma evolution dynamics} 
 
\subsubsection{Resonant enhancement of the ultracold plasma SFI and late-peak signal}

The UV-UV double-resonance spectra plotted in Figure \ref{fig:no_mm} show that, among accessible prepared $N=1$ states of Rydberg gas, the avalanche to ultracold plasma strongly selects for those with high orbital angular momentum.  The resonances detected by SFI or as an ultracold plasma late peak belong almost entirely to the single, atom-like $\ell=3$ series, $n_0f(2)$.  The only evidence for states of lower $\ell$ appears at positions where they occur as $N=1$ interlopers, that mix with neighbouring states in the $n_0f(2)$ series.  Here, they borrow lifetime conferred by admixed, non-penetrating $\ell=3$ character.  

This selectivity clearly occurs as a consequence of predissociative decay to form neutral N($^4$S) plus O($^3$P).  Among optically accessible states with $N=1$, only those with highest orbital angular momentum, $\ell=3$, present a barrier sufficient to suppress the e$^-$ - NO$^+$ ion core coupling that causes ultrafast predissociation.  Long-lived NO high-Rydberg molecules undergo Penning ionization and electron impact avalanche to form an ultracold plasma.  

Promotion of surviving, optically accessible $n_0f$ Rydberg molecule to a state of orbital angular momentum $|n_0,\ell=3>$ ($(n_0\pm 1)g$) further decreases core penetration and extends lifetime.  We observe this consequence of an increase in $\ell$ by an evident increase in the plasma yield, detected after 300 ns in the SFI signal and 20 $\mu$s later in the integrated amplitude of the plasma late peak.  Separate evidence suggests that evolution to a plasma state of arrested relaxation relies on an balance in the density of surviving Rydberg molecules and NO$^+$ ions \cite{Marroquin2020}.  

Experimental results plotted in Figure \ref{fig:mm_delay} serve to define a critical time interval of  evolution after $\omega_2$, during which mm-wave plasma enhancement occurs in relation to spectroscopically adjacent suppression.  We see here that a mm-wave field delayed to begin as little as 1 $\mu$s after $\omega_2$ acts with greatly diminished effect in promoting plasma by means of $n_0f(2) \rightarrow (n_0 \pm 1)g(2)$ resonant absorption.  

We can interpret this falling amplitude ratio as a direct consequence of $\ell$ mixing that occurs on a 1 $\mu$s timescale owing to Penning electron-Rydberg collisions.   As electron collisions drive the optically selected high-$n$ $\ell=3$ Rydberg orbital angular momentum distribution to occupy states of higher $\ell$, the advantage of a mm-wave resonant $f$ to $g$ transition disappears.  

The rate at which this mm-wave resonant intensity decreases in Figure \ref{fig:mm_delay} agrees with the timescale at which we observe a blue shift of $n_0$ Rydberg features in the time-resolved SFI spectrum signalling $\ell$-mixing in a nitric oxide Rydberg gas during the first 100 ns of  avalanche to plasma \cite{Haenel2017}.  The prominence of such lifetime lengthening effects in plasma action spectra underline the importance of Rydberg dynamics in determining plasma stability. 

\subsubsection{Resonant suppression of the ultracold plasma SFI and late-peak signal}

The right-hand frames of Figure \ref{fig:w3} show clear resonant dips in the late-peak $\omega_{\rm mm}$ spectra of selected $n_0f(2)$ Rydberg molecules.  Figure \ref{fig:QD} establishes that the positions of these resonances fit well with assignment to the series of transitions, $n_0f(2) \rightarrow (n_0 \pm 1)d(2)$.  The lineshapes of the dips in Figure \ref{fig:w3} amply overlap the energies of transitions originating from the nearly iso-energetic less-penetrating states, $n_0\ell(2)$ for $\ell \ge 4$ formed by $\ell$-mixing. 

Isolated $nd$ Rydberg states of NO predissociate on a nanosecond timescale for principal quantum numbers in this range \cite{Bixon,Vrakking1995,Remacle1998,Murgu2001}, and this dissipation must account for the fact that $\Delta n_0 = \pm 1$ transitions cause the late-peak plasma density to decrease.  A Rydberg-Rydberg transition to an $nd$ state from any originating level with $\ell \ge 3$ drives population from a long-lived state to a predissociative one.  

But, we find that promotion to a predissociatve $(n_0 \pm 1)d(2)$ state has very little effect on the $\omega_{\rm mm}$ spectrum recorded at short-time in the SFI signal.  This is not hard to explain.  Electron-collisional $\ell$-mixing, which erases the short-time advantage of $n_0 f(2) \rightarrow (n_0 \pm 1)g(2)$ excitation, acts as well to $\ell$-mix  $(n_0 \pm 1)d(2)$ states populated by mm-wave absorption in the dips during the avalanche.  Predissociatve $\ell=2$ levels make up a very small fraction of the local density of states at any higher value of $n$.  Inelastic electron-Rydberg collisions thus expose these spectroscopically selected molecules to a substantial statistical pressure that favours redistribution to long-lived, higher-$\ell$ states, erasing the resonant dip in the short-time in the $\omega_{\rm mm}$ spectrum.  

Judging from Figure \ref{fig:mm_delay}, $\ell$-mixing fails to diminish the $n_0f,g,\ell(2) \rightarrow (n_0 \pm 1)d(2)$ resonant dips observed in spectra recorded by collecting the late-peak plasma signal after 20 $\mu$s.  Clearly, electron-collisional redistribution does not compete with predissociation after a few microseconds of plasma evolution.  

However, in Figure \ref{fig:mm_delay}, SFI spectra recorded over this same interval of time show that most of the population initially in $n_0 f(2)$ or $\ell$-mixed to $n_0 \ell(2)$ spreads away from $n_0$ to occupy an broad distribution of principal quantum numbers, rising to a peak within 200 GHz of the ionization threshold.  

The late-peak resonant dips tell us that $n_0f,g,\ell(2) \rightarrow (n_0 \pm 1)d(2)$ resonant excitation operates efficiently to deplete the detected plasma signal.  We know, therefore, that the electron-Rydberg collisional $\ell$-mixing processes, that prevent these dips from appearing in the SFI-detected $\omega_{\rm mm}$ spectra, must cease in the evolving plasma.  

But, after an evolution of 1 $\mu$s or so, the long-lived $n_0 \ell(2)$ originating state of this resonance forms a very small fraction of the plasma.  Therefore, some mechanism must exist to refill the hole in the distribution inevitably burned by excitation and predissociation.  We examine this question in the next section.

\subsubsection{Local Rydberg resonance and the time-evolution of global plasma suppression}

Collision with Penning and early avalanche electrons scrambles the orbital angular momentum of $n_0f(2)$ Rydberg molecules on a 100 ns timescale.  All but high-$\ell$ molecules predissociate.   Electron-collisional redistribution of molecules to low-$\ell$ evidently ceases, and the plasma evolves to form an arrested state that includes a small fraction of surviving $n_0\ell (2)$ molecules.  As demonstrated in Figure \ref{fig:mm_delay}, this reaches its completion on the $\sim 1 ~\mu$s timescale with which the positive-going structure in the mm-wave spectrum disappears with a short-time delay in the application of $\omega_{\rm mm}$ after $\omega_2$.  

The dips behave differently.  On a timescale of at least 2 $\mu$s, undiminished, broad, negative-going mm-wave resonances persist in the regions of the $n_0f,g,\ell(2) - (n_0 + 1)d(2)$ and $n_0f,g,\ell(2) - (n_0 - 1)d(2)$ transitions.   These dips tell us that a depletion of the Rydberg population at any time decreases the total plasma signal detected as the late peak.  This persistent decrease occurs despite evidence in the SFI spectrum showing that the $n_0$ Rydberg molecules account for only a very small fraction of the plasma density.  Thus, $n_0$-Rydberg depletion must be accompanied by hole-filling.  As a sustained effect, the plasma loses as much as three-quarters of its population.  

This begs the question of repopulation.  What dynamical process refills the hole in the $n_0$-Rydberg population, and enables resonantly driven predissociation to continue nearly to completion?  We cannot explain $n_0$ hole filling as an electron-collision driven process.  The $\ell$-mixing that accompanies $n$-changing collisions of this nature would readily cycle the Rydberg population through low $\ell$ and cause plasma decay on a microsecond timescale in the {\emph {absence}} of a resonant mm-wave field.  

In a recent effort to model properties of the NO molecular ultracold plasma responsible for its evolution to a state of arrested relaxation, we have argued that dipole-dipole exchange interactions define a localized network of coupled Rydberg levels.  This network appears as an end state of a Rydberg gas avalanche, bifurcation and quench that forms an ultracold, strongly correlated ensemble of NO$^+$ ions, electrons and high-Rydberg molecules.  During this process, low-$\ell$ molecules predissociate.  This purifies the plasma, constraining its Rydberg component to molecules in states of high-$\ell$.  In the absence of mm-wave radiation, this system remains stable with respect to predissociation.  

We crudely assume that the particles in this nearly static post-quench regime form a system of interacting two-level dipolar pseudospins, in which each dipole randomly samples from a complete central-field basis to define an ensemble state that solves the XXZ Hamiltonian:  
\begin{equation} \label{XY-Ising}
H_{\rm eff} = \sum_i \epsilon_i \hat{S}^z_i + \sum_{i,j} J_{ij} (\hat{S}^+_i \hat{S}^-_j + h.c.) + \sum_{i,j} U_{ij} \hat{S}^z_i \hat{S}^z_j,
\end{equation}
where the term $\sum_i \epsilon_i \hat{S}^z_i$ defines a Gaussian random local field that spans a width, $W$, measured by the distribution of binding energies observed in the SFI spectrum.  Dipole-dipole matrix elements describe the resonant transfer of excitation between $i$ and $j$,  where $J_{ij} = {t_{ij}}/{ {r}_{ij}^3}$ determines the amplitudes of these spin flip-flop interactions \cite{agranovich,Gorshkov,XiangKrems}.  To give an example, a $t_{ij}$ ($C_3$) of 3230 MHz $\mu$m$^3$ measured by Ga\"etan et al. \cite{Gaetan} for the $58d_{3/2} + 58d_{3/2} \leftrightarrow 60p_{1/2} + 56f_{3/2}$ interaction in $^{87}$Rb scales as $n^2$ to yield a dipole-dipole coupling energy of 2 GHz for a density 0.1 $\mu$m$^{-3}$.  The third term refers to Ising interactions, where $U_{ij} = {D_{ij}}/{r}_{ij}^6$ and $D_{ij} = {t_{ij}^2\widetilde{J}}/{W^2}$.  These assume an increasing importance for larger system sizes.  

Under such conditions, the introduction of a depletion specific to one level serves to connect all levels to a bath of free N($^4$S) plus O($^3$P) neutral atoms.  We assume no back interaction of these product atoms with the ions, electrons and Rydberg molecules in the plasma.  The mm-wave transition simply opens a channel that dissipates energy and mass by the irreversible conversion of ions and electrons to escaping neutral atoms.  As the vast network of pairwise interactions evolves to fill the hole burned by the mm-wave transition, it transforms the entire ensemble into an open quantum system.  In principle, a field tuned to any $n \ell (2) \rightarrow (n + 1)d(2)$ resonance would produce the same dissipation, regardless of the particular value of $n_0$ initially selected.  

Note that the baseline stability of these $n \ell (2)$ states coupled with the distinct resonant response of the plasma to mm-wave radiation points to a signature spectroscopic characteristic of the residual Rydberg population in the arrested state of the nitric oxide molecular ultracold plasma.  Long-lived Rydberg states of NO derive their stability from an ${\rm e^- -NO^+}$ recombination barrier owing to high orbital angular momentum.  This core isolation confines energy level positions to those of near-zero quantum defect.  Predissociation pressure thus gives the Rydberg residue an  energy landscape confined to the periodic structure of the $n \ell (2)$ series.  This small portion of the plasma binds electrons in bands of energy, $-R/n^2$, with respect to rotational quantum states of NO$^+$, where $R$ is the Rydberg constant of NO and $n$ is an integer that falls, for present purposes, in the range from 38 to 80.   

The durability of this structure plays an important role in plasma lifetime, because, as the experiment shows, the system responds to a local predissociative instability by repopulation or hole filling that eventually depletes the entire plasma.  The stability of the plasma speaks to the evident durability of this banded structure, and proves that electron-collisional $\ell$-mixing ceases as the plasma enters the arrested state.  

\subsection{Lineshapes of Rydberg-Rydberg transitions in the nitric oxide molecular ultracold plasma}

\subsubsection{Functional form of fits to $n_0f(2) \rightarrow (n_0\pm 1)g(2)$ and $n_0f,g,\ell(2) \rightarrow (n_0\pm 1)d(2)$ resonances}

The positive-going $n_0f(2) \rightarrow (n_0\pm 1)g(2)$ resonances highlighted in Figure \ref{fig:w3} display distinctive $\sim 1$ GHz linewidths, both as features measured after 300 ns in the total SFI signal and as structure detected 20 $\mu$s later in the amplitude of the late peak.  We cannot readily explain these linewidths based on the properties of isolated molecules.  These $f$ and $g$ Rydberg states have intrinsically long lifetimes.  Rydberg-Rydberg transitions do occur with very large dipole transition moments, but the present measurements take place in a range of linear absorption that only approaches saturation. Figure \ref{fig:power}, establishes that even the most intense mm-wave fields fall short of detectable power broadening.  

 All peaks show a small but very evident degree of asymmetry.  These asymmetric lineshapes conform well with conventional anti-resonant Fano profiles, described by a function:
  \begin{equation}
\sigma(\epsilon)=\frac{(q+\epsilon)^2}{1+\epsilon^2},
\label{eqn:Fano1}
\end{equation}
that expresses the absorption cross section, $\sigma(\epsilon)$, in terms of an asymmetry parameter, $q$, and detuning, $\epsilon$, defined by:
\begin{equation}{}
 \epsilon=\frac{\omega_{\rm mm}-\omega_0}{\Gamma/2}
 \end{equation}
in which $\omega_0$ represents the resonant frequency of the transition and $\Gamma$ describes its linewidth.  
 
 Table \ref{tab:fano-g} summarizes parameters that pertain to all of the $n_0f(2) \rightarrow (n_0 \pm 1)g(2)$ resonances fit in Figure \ref{fig:w3}
 
\begin{table}[h!]
	\centering
	\caption{Fano parameters for $n_0f(2)$ to $(n_0 \pm 1)g(2)$ Rydberg-Rydberg transitions }
	\label{tab:fano-g}
		\resizebox{.45\textwidth}{!}{%
		\begin{tabular}{crcrrrr}
	\toprule
	\multirow{2}{*}{$n_0f(2)$}   &   &   \multicolumn{3}{c}{$n_0f(2)\rightarrow (n \pm 1)g(2)$ }   &  \\
	&    \multicolumn{1}{c} { $\omega_0$ }   &  $\Gamma$   &   $q$   &   \multicolumn{1}{c} { \hspace{10 pt} $\omega_0$ } &  \multicolumn{1}{c} { \hspace{5 pt}  $\Gamma$   }   &   $q$   \\
	 & (GHz) &    (GHz)  &      &    (GHz)&     (GHz)             \\
   			\midrule  
$39f(2)$ & 108.84  & 0.8  & 20  & 113.22   & 0.8  & -20 \\
$40f(2)$ & 101.18   & 0.9  & 10  & 104.91  & 0.9  & -10 \\
$41f(2)$ & 94.08   & 0.8 & 20  & 97.25   & 0.8 & -20 \\
$42f(2)$ & 87.61   & 1.0    & 15  & 90.22   & 1.0    & -15 \\
$43f(2)$ & 81.86   & 0.9  & -15 & 83.92   & 0.9  & 15  \\
$44f(2)$ & 76.33   & 1.0    & 12  & 78.26    & 1.0    & -12 \\
$45f(2)$ & 71.25   & 0.8  & 20  & 73.23   & 0.8  & -20 \\			
\bottomrule
		\end{tabular}%
		}
\end{table}

Notice that most of the lower-energy $n_0f(2) \rightarrow (n_0 + 1)g(2)$ transitions, described by positive $q$, shade to the blue, while lineshapes for $n_0f(2) \rightarrow (n_0 - 1)g(2)$ transitions best fit to negative $q$, and shade to the red, except for the resonances that originate from $43f(2)$ for which this asymmetry reverses.  Relatively large values of $q$ cause lineshapes in all of these cases to approach Lorentzians.  
 
Resonances assigned to $n_0\ell(3) \rightarrow (n_0 \pm 1)d(2)$ appear as dips in $\omega_{\rm mm}$ late-peak excitation spectra with broader linewidth and greater asymmetry.  
  \begin{table}[h!]
	\centering
	\caption{Fano parameters for $n_0\ell(2)$ to $(n_0 \pm 1)d(2)$ Rydberg-Rydberg transitions}
	\label{tab:fano,g-d}
	\resizebox{.46\textwidth}{!}{%
		\begin{tabular}{crcrrcr}
			\toprule
	\multirow{2}{*}{$n_0f(2)$}   &  &   $n_0\ell(2)$ & $\rightarrow$ & $(n \pm 1)d(2)$   &  & \\
	&    \multicolumn{1}{c} { $\omega_0$ }   &  $\Gamma$   &   $q$   &   \multicolumn{1}{c} { \hspace{16 pt} $\omega_0$ } &  $\Gamma$      &   $q$   \\
		 & (GHz) &    (GHz)  &      &    (GHz)  &     (GHz)             \\
			\midrule  
$39f(2)$ & 116.42 &  2 & 10 & 105.78 &  2 & -10 \\
$40f(2)$ & 108.10 &  2 & -2 & 98.13  &  2 & 2   \\
$41f(2)$ & 100.12 & 2 & 10 & 91.32  &  2 & 10  \\
$42f(2)$ & 93.07  & 3 & -4 & 85.11  &  3 & -4  \\
$43f(2)$ & 86.10   &  2 & 4  & 79.04  &  2 & 4   \\
$44f(2)$ & 80.34  &  2 & 4  & 74.24  &  2 & -4  \\
$45f(2)$ & 75.65  &  2 & -5 & 69.42  & 2 & 5    \\ 
		
\bottomrule
		\end{tabular}%
		}
\end{table}

Table \ref{tab:fano,g-d} collects the parameters that describe the Fano parameters for all the dips fit to lineshape functions plotted in Figure \ref{fig:w3}.  

 \subsubsection{The dynamics that govern Rydberg-Rydberg anti-resonant lineshapes} 
 
 A system of dipoles exhibits an electromagnetic spectrum that reflects its time-dependent polarization when stimulated by an infinitesimally short pulse.  A delta-function response implies a spectroscopic continuum.  The exponential response of a discrete excited state produces a Lorentzian lineshape of some characteristic width, $\Gamma$.  The presence of both - a discrete state embedded in a dipole continuum to which it couples - gives rise to a characteristic Fano lineshape \cite{Greene}.  Spectroscopically, the driven system coherently adds a discrete-state Lorentzian to the continuum response, shifted in phase to form the asymmetric, anti-resonant profile described by Eq (\ref{eqn:Fano1}), in which the phase shift, transformed to the frequency domain, determines $q$.  
 
 Atomic and molecular spectroscopists have applied this formalism extensively to describe configuration interaction as it affects the autoionization dynamics of neutral excited states embedded in ionization continua \cite{Fano,Gallagherfano,madden,rost,Klinker,minns,bryant2,viteri}.  Wavefunction interference also provides a useful framework for interpreting bound-quasi-continuum interactions in complex materials, including quantum dots \cite{kobayashi} and plasmonic nanostructures \cite{miroshnichenko,Lukyanchuk,Liu}.  The concept of Fano resonance has proven useful in describing how compact localized states scale the effect of disorder in flat-band lattice systems \cite{Sergej,Leykam}, and reflect the degree to which central-site interactions mix the wavefunctions in a surrounding, Anderson localized (AL) chain \cite{Hetterich}.

A dynamics of avalanche and bifurcation appear to quench the present nitric oxide molecular ultracold plasma to a highly disordered, strongly coupled state of anomalously slow relaxation \cite{Haenel2017,review}.  This evolution, directly monitored by experiment, creates conditions in the system that approach the threshold for many-body localization (MBL) \cite{gopalakrishnan2020dynamics}.  We have taken information from experiment to rationalize the behaviour of the NO ultracold plasma in terms of a minimal phenomenological model of randomly interacting dipoles of random energies that gives rise to a system in or near a state of MBL.

In addition to the question of MBL as a enduring phase in various dimensions, theorists have raised important questions about how a localized system responds to contact with a bath or a rare thermal region \cite{Mott,nandkishoreSpec,NandkishorePRL,abanin2019colloquium,agarwal}.  Such questions of imperfect isolation are of great concern to experimentalists \cite{rubio2019many}.  Generally speaking, delocalizing interactions act to mobilize transport and mix localized states with consequences for the statistics of energy level spacings. 

The particular simple limit of a single perturbing level acting with a coupling strength, $m$ on the quasi-continuum of an Anderson localized (AL) bath with disorder width, $W$, causes a configuration interaction that maps onto the Fano resonance problem \cite{Hetterich}.  The ratio, $\mu = |m/W|$ determines a leading-order self-energy correction that sets the position and spectral weight of the resonance.  In the limit of weak coupling, $|m/W| \ll 1$, a bright state retains its spectral weight and position, and its interaction with the AL bath fails to cause a Poisson energy distribution of eigenvalues to acquire repulsion.  A coupling strength that substantially exceeds the width of the disorder,  $|m/W| \gg 1$, shifts the resonance by more than $\mu W$, and places most of the spectral weight in a $[-W,W]$ continuum regime of repelling levels and ballistic entanglement.   

To draw a crude analogy, let us consider the very few $n_0f(2)$ Rydberg molecules remaining in the nitric oxide ultracold plasma after 1 $\mu$s as points of origin for mm-wave transitions to particular selected bright states in a disordered dipolar system that spans an energy width $[-W,W]$, which extends over a measured interval of about 500 GHz.  This disorder width substantially exceeds the 1 GHz average dipole-dipole coupling strength in the arrested state of the quenched plasma.  In this bright-state picture, each of the pair of $n_0f(2) \rightarrow (n_0 \pm 1)g(2)$ Rydberg-Rydberg transitions in Figure \ref{fig:w3} yields a Fano lineshape consistent with a weak coupling regime, in which, the coupling of the bright state determines the linewidth, $\Gamma \approx m$ \cite{goldschmidt2016anomalous}, the spectral weight favours the discrete-discrete transition, $q > 10$, and dipole-dipole interactions fail to drive level repulsions.  

From this perspective, the spectrum of transitions from high-$\ell$ to low-$\ell$ appears qualitatively different.  Time-resolved experiments show that $n_0d(2)$ states predissociate on a nanosecond timescale \cite{Bixon,Vrakking1995,Remacle1998,Murgu2001}.. This is too slow to explain the observed values of $\Gamma = 2$ GHz as an uncertainty bandwidth.  However, a spectroscopically accessed state that connects the bath of NO dipoles via fast predissociation to a continuum of free ${\rm N(^4S) + O(^3P)}$ atoms could alter the coupling and level statistics of the bath in such a way as to affect the lineshape of the Rydberg-Rydberg transition.  In particular, stronger coupling ought to produce a broader linewidth and an increase in the spectral weight of the continuum, resulting in a broader, more asymmetric lineshapes, as measured experimentally by dips in the late peak plasma signal.  

That these resonances appear as dips, and that the associated depletion far outweighs the residual fraction of $n_0$ Rydberg molecules speaks to a transport property that must arise in the ultracold plasma to fill the hole created by $n_0 f,g,\ell(2) \rightarrow (n_0 \pm 1)d(2)$ excitation and predissociation.  Whatever the dynamics that drive this hole filling, they appear to require $(n_0 \pm 1)d(2)$ excitation to operate.  

Thus, the subset of predissociative states created by mm-wave excitation must act differently on the dipoles in the arrested ultracold plasma to increase the range of state-changing interactions.  The $(n_0 \pm 1)d(2)$ states differ by virtue of their nanosecond rate of predissociation.  Predissociation bridges the  $nd$ excited state, and by extension, the entire arrested system to the continuum of free ${\rm N(^4S) + O(^3P)}$ atoms.  

By forging this connection to a thermal continuum via predissociation, mm-wave driving in effect adds a dissipative term to, Eq (\ref{XY-Ising}), the XXZ Hamiltonian that describes dipolar coupling.  Even the simple limit of a minimal subsystem of localized pseudospins linked to a dissipative bright state presents a significant challenge.  Still, one can imagine intriguing realms of tuned bright-state-subsystem coupling and predissociation width.  Despite the limitation of greatly simplifying approximations, such calculations could offer the promise intriguing links to spectroscopic measures of localization instability.  

\section{Conclusions}

Double-resonant excitation of NO in a supersonic molecular beam creates a state-selected Rydberg gas that evolves to form an ultracold plasma.  The efficiency of this process depends sensitively on  the dissociative lifetime of the initially selected Rydberg quantum state.  Among the exclusively $N=1$ levels accessible via intermediate resonance with A $^2\Sigma^+$ $N'=0$ (where $N$ refers to total angular momentum neglecting spin), only states in the $\ell=3$ series converging to $N^+=2$ $\left(n_0f(2) \right)$ evolve to a long-lived plasma state.  Higher-$\ell$, $n_0g(2)$ states live an order of magnitude longer,  Thus, mm-wave radiation tuned to $n_0f(2) \rightarrow (n_0 \pm 1)g(2)$ transitions substantially increases the density of plasma that survives the 300 ns of avalanche and persists to form a late-peak electron signal after 20 $\mu$s or more. 
 
The mm-wave excitation spectrum also exhibits transitions from states, $n_0\ell(2)$, to $(n_0 \pm 1)d(2)$ that flank each pair of non-penetrating resonances.  Here, $\ell$ refers to orbital angular momenta $\ell \ge 3$, for which quantum defects cluster near zero. Photoproduced $nd$ Rydberg states of NO predissociate with nanosecond lifetimes, and these $(n_0 \pm 1)d(2)$ resonances appear as small dips in $\omega_{\rm mm}$ excitation spectra, sampled under avalanche conditions by the SFI signal at 300 ns.  By contrast, $nd$ predissociation decreases the plasma signal to near-zero in resonance dips sampled in the late-peak after a long-time evolution of 20 $\mu$s.   

At early times, the SFI spectrum shows that the evolving plasma retains a substantial proportion of $n_0f(2)$ Rydberg molecules.  On a time-scale of about 1 ns, $\ell$-mixing electron-Rydberg collisions drive this population to states of higher orbital angular momentum, and the  $n_0f(2) \rightarrow (n_0 \pm 1)g(2)$ resonances disappear.  

These transformations have no effect on the resonance condition for $n_0\ell(2)$ to $(n_0 \pm 1)d(2)$ dips, and delaying the onset of the mm-wave field by as much as 2 $\mu$s does not diminish the amplitude of the dip detected in the late-peak signal.  

However, despite an ample density of resonant molecules, $n_0\ell(2) \rightarrow (n_0 \pm 1)d(2)$ excitation spectra detected in the short time SFI signal show very little dip amplitude.  We can understand this as an effect of $\ell$-mixing electron-Rydberg collisions that compete on the nanosecond timescale of low-$\ell$ predissociation to scatter $nd$-Rydberg molecules to states of higher orbital angular momentum.  

The $nd$-Rydberg states populated later dissociate as the plasma evolves, forming a resonant dip in the late-peak signal, in some cases nearly to zero.  SFI spectra show that this arrested state contains only a very small residual population of $n_0\ell(2)$ molecules.  

Therefore, $n_0\ell(2) \rightarrow (n_0 \pm 1)d(2)$ excitation must instantly burn a hole in the distribution of Rydberg molecules.  Predissociation cannot proceed to a state of deep depletion without some mechanism to repopulate the originating state.  

The fact that $(n_0 \pm 1)d(2)$ molecules predissociate with effectiveness tells us that stabilizing electron-Rydberg collisions cease in the arrested state of the plasma.  Thus, some form of collisionless energy redistribution must occur for $n_0\ell(2) \rightarrow (n_0 \pm 1)d(2)$ excitation and predissociation to persist.  We see no evidence for energy redistribution in the absence of mm-wave transitions that populate predissociative $nd$ states.  This suggests that predissociation itself plays a role in the hole-filling required to sustain it.  

We suggest the model of a spectroscopically bright state that bridges the quenched ultracold plasma to a neutral continuum offers a possible way to think about this process.  The plasma exists in an arrested state, consisting of NO$^+$ ions, electrons and high-$\ell$ Rydberg molecules in quasi-equilibrium - perhaps under conditions of localization.  This plasma reaches its quenched state depleted in $nd$ molecules by predissociation.

Promotion of a single subset of $n_0\ell(2)$ molecules in this ensemble to a bright state broadened by predissociative coupling to the ${\rm{N(^4S) + O(^3P)}}$ continuum bridges this continuum to the ensemble of localized dipoles.  This coupling mixes states in this localized ensemble, creating a flux of molecules that can absorb mm-wave radiation resonantly within the bandwidth of the $n_0\ell(2) \rightarrow (n_0 \pm 1)d(2)$ transitions.  

The evidence for this coupling appears in a broadened linewidth and spectral weight transferred to the continuum that gives rise to anti-resonant lineshape.  This spectroscopic manifestation offers an intriguing possible point of contact between theory and quantum state resolved experiment.  

\begin{acknowledgements}

This work was supported by the US Air Force Office of Scientific Research (Grant No. FA9550-12-1-0239), together with the Natural Sciences and Engineering research Council of Canada (NSERC), the Canada Foundation for Innovation (CFI) and the British Columbia Knowledge Development Fund (BCKDF).  JS gratefully acknowledges support from the National Science Foundation (NSF) Materials Research Science and Engineering Centers (MRSEC) program through Columbia University in the Center for Precision Assembly of Superstratic and Superatomic Solids under Grant No. DMR-1420634.

\end{acknowledgements}

\bibliography{ref}

\end{document}